\documentclass[preprint,12pt,3p]{elsarticle}

\usepackage{amssymb}
\usepackage{amsmath}
\usepackage{url}

\usepackage{ragged2e}
\usepackage{color}

\journal{Neuroimage}

\begin{document}

\begin{frontmatter}

\title{Cortical surface registration using unsupervised learning}

\address[label1]{A.A. Martinos Center for Biomedical Imaging, Massachusetts General Hospital, Harvard Medical School, USA}
\address[label2]{Computer Science and Artificial Intelligence Laboratory, Massachusetts Institute of Technology, USA}

\author[label1]{Jieyu Cheng}

\author[label1,label2]{Adrian V. Dalca}

\author[label1,label2]{Bruce Fischl \corref{cor1}}

\author[label1]{Lilla Z\"ollei \corref{cor1}}
\ead{lzollei@nmr.mgh.harvard.edu}

\author{\\ for the Alzheimer's Disease Neuroimaging Initiative \corref{cor2}}

\cortext[cor1]{These authors share senior authorship}
\cortext[cor2]{Data used in preparation of this article were obtained from the Alzheimer’s Disease Neuroimaging Initiative (ADNI) database (adni.loni.usc.edu). As such, the investigators within the ADNI contributed to the design and implementation of ADNI and/or provided data
but did not participate in analysis or writing of this report. A complete listing of ADNI investigators can be found at: \url{http://adni.loni.usc.edu/wp-content/uploads/how_to_apply/ADNI_Acknowledgement_List.pdf}}

\begin{abstract}
Non-rigid cortical registration is an important and challenging task due to the geometric complexity of the human cortex and the high degree of  inter-subject variability. A conventional solution is to use a spherical representation of surface properties and perform registration by aligning cortical folding patterns in that space. This strategy produces accurate spatial alignment, but often requires high computational cost. Recently, convolutional neural networks (CNNs) have demonstrated the potential to dramatically speed up volumetric registration. However, due to distortions introduced by projecting a sphere to a 2D plane, a direct application of recent learning-based methods to surfaces yields poor results. In this study, we present SphereMorph, a diffeomorphic registration framework for cortical surfaces using deep networks that addresses these issues. SphereMorph uses a UNet-style network associated with a spherical kernel to learn the displacement field and warps the sphere using a modified spatial transformer layer. We propose a resampling weight in computing the data fitting loss to account for distortions introduced by polar projection, and demonstrate the performance of our proposed method on two tasks, including cortical parcellation and group-wise functional area alignment. The experiments show that the proposed SphereMorph is capable of modeling the geometric registration problem in a CNN framework and demonstrate superior registration accuracy and computational efficiency. The source code of SphereMorph will be released to the public upon acceptance of this manuscript at \url{https://github.com/voxelmorph/spheremorph}
\end{abstract}

\begin{keyword}
cortical surface registration, deep learning, unsupervised learning, SphereMorph, subject-to-atlas registration
\end{keyword}

\end{frontmatter}


\section{Introduction}
\label{sec:Intro}
Non-rigid shape registration is an important area of research in medical imaging, in particular for establishing cross-subject spatial correspondence in the cerebral cortex.  This type of spatial alignment has been shown to improve the statistical power of group functional MRI (fMRI) analysis~\cite{van2004integration,frost2012measuring} resulting from the improved correspondence of functional areas. Due to the geometric complexity of the cortex and the large variability between individuals, cortical surface registration remains a challenging task. Inter-subject surface alignment is commonly driven by geometric features that describe measures of cortical shape (folding), such as sulcal depth or local curvature~\cite{fischl1999fsreg,yeo2010sphericaldemons,conroy2013inter,tardif2015multi}.

A widely used cortical surface registration approach is to map the surface onto the unit sphere in order to perform computations in this canonical domain. Existing efforts are mainly focused on the adaptation of registration algorithms in the Euclidean space~\cite{fischl1999fsreg,yeo2010sphericaldemons,robinson2014msm}. These aim to optimize a similarity metric between the target and the deformed source volumes, regularized by various energies~\cite{sotiras2013deformable}. FreeSurfer~\cite{fischl1999fsreg} registers an individual surface to a probabilistic atlas computed from a representative set of subjects by minimizing the squared difference between the average convexity across subjects and that of the individual, weighted by the inverse variance of the convexity across subjects. The Multimodal Surface Matching tool~\cite{robinson2014msm} uses a similarity between the input and reference mesh features in a coarse-to-fine manner. The deformation is driven by aligning local patches around control points, i.e. vertices of a low resolution mesh, and then propagated to the high resolution input mesh via interpolation. Spherical Demons~\cite{yeo2010sphericaldemons} modifies the classical Demons method~\cite{thirion1998image} using velocity vectors tangent to the sphere. The two-step optimization of classical Demons also holds for the spherical case in which the second step handles the deformation regularization by spherical thin plate spline interpolation. To encourage desirable mathematical properties such as invertibility, diffeomorphic transforms have seen extensive methodological development, yielding state-of-the-art tools~\cite{ashburner2007DARTEL,zhang2017frequency}. Unfortunately, since these methods solve an optimization problem for each image pair, they often exhibit long execution times. 

High computational costs have led to an increase in the popularity of supervised~\cite{krebs2017robust,sokooti2017nonrigid,yang2016pred} and unsupervised~\cite{balakrishnan2019tmi,dalca2018miccai,dalca2019varreg,niethammer2019metric,wang2015predict} learning-based registration algorithms. Considering the difficulty of establishing ground truth spatial correspondences, supervised methods require predictions from existing algorithms~\cite{yang2016pred}, simulations~\cite{sokooti2017nonrigid}, or both~\cite{krebs2017robust}. In contrast, unsupervised methods make use of Spatial Transformer Networks (STN)~\cite{jaderberg2015stn} to warp the moving image in a differentiable way, enabling end-to-end training~\cite{balakrishnan2019tmi,dalca2019varreg,jason2016back,krebs2019learning,de2019deep,niethammer2019metric}. Some unsupervised methods~\cite{balakrishnan2019tmi,dalca2019varreg,krebs2019learning} model a stationary velocity field as latent variables representing deformations in a generative probabilistic model. They use a scaling and squaring layer~\cite{arsigny2006log} for the Lie group exponentiation of the velocity field to generate diffeomorphic transforms, thus guaranteeing topology preservation. These methods have demonstrated high quality performance in registering various types of medical images. Therefore, we build on these concepts when working with surfaces. A more complex model~\cite{niethammer2019metric} utilizes a vector momentum-parameterized stationary velocity field (vSVF) and jointly optimizes the local regularizer parameterized by a deep network and the registration parameters of the vSVF model for location-varying regularization. In addition to velocity field registration models, sparse learning is also explored for image-template key point matching~\cite{wang2015predict} and a subsequent interpolation to a dense deformation field via radial basis functions. However, the performance is dependent on the key point selection.

Recent studies have developed geometric convolutional neural networks (CNN)~\cite{su2017learning,cohen2018spherical,coors2018spherenet,seong2018geometric,jiang2019sphericalunet,zhao2019sphereunet} that operate on a spherical manifold to solve classification and detection tasks. To address the distortions introduced by projecting signals to a planar image, regular convolutions with increased kernel sizes near polar regions have been utilized~\cite{su2017learning}. Spherical CNNs encode rotational instead of translational equivariance into the network in Euclidean space to solve classification problems~\cite{cohen2018spherical}. In SphereNet, the convolution kernel on the sphere has been approximated by encoding the vertex neighborhood information on 2D tangent planes, which enables adapting existing CNN architectures to the omnidirectional setup for object detection and classification tasks~\cite{coors2018spherenet}. Geometric CNNs (gCNN)~\cite{seong2018geometric} also deal with convolution and pooling operations of a CNN on a mesh surface. However, they have so far only been tested on sex classification, using cortical thickness images. A convolutional kernel discretized by an unstructured mesh was recently proposed and evaluated on spherical MNIST classification and 3D object detection tasks~\cite{jiang2019sphericalunet}. Spherical U-Net~\cite{zhao2019sphereunet} proposes a novel Direct Neighbor convolutional kernel based on expansion and contraction process of icosahedron and defines corresponding convolution and pooling operations. Graph convolutions~\cite{gopinath2019graph,gopinath2019adaptive} have been utilized for brain surface data in aligned spectral domains to learn the node-wise prediction~\cite{gopinath2019graph}, e.g. cortex parcellation and global subject-wise information~\cite{gopinath2019adaptive}, e.g. disease classification or age regression. FastSurfer~\cite{henschel2020fastsurfer} introduces a full alternative pipeline for FreeSurfer and omits nonlinear surface-atlas registration via fast spherical mapping that quickly maps the volumetric parcellation to cortex using Laplace Eigenfunctions. Most existing work, however, focuses on the construction of spherical convolutional kernels and, to the best of our knowledge, neural networks have not yet been extended to surface registration. Compared with classification or detection tasks, besides convolution and pooling operations on spheres, a learning-based registration method should address local deformations defined on spheres. However, existing spatial transformation networks~\cite{polarSTN2017,tai2019equivariant} for spheres only address global deformations, which are not suitable for accommodating nonlinear deformation fields.

In this paper, we propose a diffeomorphic framework combining a generative model for surfaces with CNNs to register individual cortical surfaces to an atlas space. This framework adapts conventional VoxelMorph for registering Euclidean images~\cite{dalca2019varreg} to spherical manifolds. In order to address the limitations of 2D planar projection, we construct a weighted neighborhood graph defined on 2D grids, which accounts for the non-uniform metric tensor of the spherical representation to encode a stationary velocity field. Considering that the 2D projection operation samples the arc-length for each latitude to the same number of points, we also take sampling distortion into account at different latitudes in the likelihood model. We quantify the performance of our framework through two applications: the generation of cortical parcellations and the alignment of functional activations. The experimental results demonstrate that our framework yields better registration accuracy to state-of-the-art classical methods at a significantly reduced computational cost, and more accurate results compared to current learning-based methods. 

Our contributions can be summarized as:
\begin{enumerate}
\item We propose a learning-based framework for spherical surface registration, which provides accurate and efficient performance compared to conventional registration work;
\item We derive an Maximum a Posteriori (MAP) solution for deformation fields in the spherical domain by correcting distortions from planar projection;
\item Different from existing spherical networks, the registration results from our proposed method aid the cortex parcellation as well as various group analyses (i.e. local thickness, functional activation etc.) while other spherical networks only learn one or multiple preset measures;
\item We explore the use of different features for functional alignment.
\end{enumerate}

The remainder of this paper is organized as follows. We first introduce the cortical registration problem, review the conventional VoxelMorph~\cite{dalca2019varreg} framework, and propose our method and network structure. We then describe two evaluation experiments including cortical parcellation and functional alignment, and show results for these experiments. Finally, we present our discussions and conclusions.

\section{Methods}
\label{sec::method}

Numerous studies using FreeSurfer have demonstrated its efficacy in spherical-based cortical registration. We build on ideas for our surface representation from the FreeSurfer spherical registration~\cite{fischl1999fsreg} and model the unsupervised learning structure for the registration field following VoxelMorph~\cite{dalca2019varreg}. 

\subsection{Registration problem definition}


In the FreeSurfer spherical registration pipeline~\cite{fischl1999fsreg}, surface geometry is encoded as a convexity attribute at each mesh vertex and the representation of the atlas surface is computed from a group of adult subjects. In order to register the surfaces of an individual to the atlas space, first a white matter mesh is generated and mapped to the unit sphere by minimizing metric distortion~\cite{fischl1999fsreg}. Next, an optimal rotational alignment is computed by global search over two rotation angles on the sphere. Then a 2D canonical warp is computed to align the subject's convexity pattern with that of the mean pattern encoded in the atlas, by minimizing the mean squared difference, weighted by the inverse of the atlas variance. Our goal is to compute this canonical warp using a CNN framework.

Let $S_x$ be the unit sphere and $I_x$ the corresponding scalar field over the sphere (e.g. sulcal depth or curvature) projected into 2D longitude/latitude parameterization. Let $I_a$ be the atlas mean image as defined in FreeSurfer, and $M$ and $N$ be the number of image rows and columns. Let $\Sigma_a = diag(\sigma_1^2, \sigma_2^2 ... \sigma_{MN}^2)$ be a diagonal matrix where each diagonal element denotes the variability of the corresponding feature at a particular vertex as defined in the FreeSurfer atlas variance image. The goal is to find the spatial transformation $\Phi:S^2 \rightarrow S^2$ given $I_x$ and $I_a$ that maximizes the a posteriori probability of the transform assuming certain smoothness priors on the warp. 

\subsection{VoxelMorph}

We assume a diffeomorphic deformation based on a stationary velocity field $v$, denoted as $\Phi_v$, and adapt a generative probabilistic model following VoxelMorph~\cite{dalca2019varreg}. VoxelMorph uses Maximum a Posteriori (MAP) estimation to obtain the most likely velocity field $v^*$ at each voxel/pixel given a pair of images. VoxelMorph models the prior probability of $v$ as a zero-mean multivariate normal distribution, $p(v)= \mathcal N(v; 0,\Sigma_v)$, where $\Sigma_v$ is the covariance matrix. An individual image can be estimated by warping the FreeSurfer atlas, thus we model the warped image $I_x \circ \Phi_v$ as $p(I_x|v;I_a) = \mathcal N(I_x \circ \Phi_v; I_a, \Sigma_a)$, where $\Phi_v$ is the inverse transformation of the atlas warping to the individual. The aim is to maximize the posterior probability $p(v|I_x;I_a) = \frac {p(v)p(I_x|v;I_a)}{\int_v p(v,I_x;I_a)dv}$, where the marginalization over $v$ is intractable. In this case, a variational approximation $q_{\psi}(v|I_x;I_a)$ is adopted with parameters $\psi$, by minimizing its dissimilarity, Kullback--Leibler (KL) divergence, with the true posterior probability. For simplicity, the approximate posterior $q_{\psi}(v|I_x;I_a)$ is restricted to a multivariate normal distribution $\mathcal N(\mu_{v|I_x;I_a},\Sigma_{v|I_x;I_a})$ where $\mu_{v|I_x;I_a}$ and a diagonal $\Sigma_{v|I_x;I_a}$ are functions estimated with a U-Net core~\cite{ronneberger2015u}, as shown in Fig.~\ref{fig:workflow}. 

Using the above assumptions, maximizing the posterior probability can be approximated by minimizing the following loss: 
\begin{equation}
    L(\psi;I_x,I_a) = -E_q[\log p(I_x|v;I_a)] + KL[q_\psi (v|I_x;I_a)||p(v)],
    \label{eq:loss1}
\end{equation}
%

\noindent where $E_q[\cdot]$ operates the expectation computation given the distribution $q$ and  $q$ is short for $q_\psi (v|I_x;I_a)$. The first term describes the reconstruction loss and the second term is a KL divergence term, encouraging the estimated posterior probability $q_\psi (v|I_x;I_a)$ to be close to the prior $p(v)$. 
VoxelMorph encourages the smoothness of the velocity field $v$ by setting $\Sigma_v^{-1} = \Lambda_v = \lambda L$, where the parameter $\lambda$ controls the scale of the velocity field and $L$ is the graph Laplacian matrix defined on the Euclidean grid. $L$ is computed as $L = (D-A)$, where $A$ is the neighborhood adjacency matrix and $D$ is the graph degree matrix. Thus, Eq.~\eqref{eq:loss1} can be rewritten as:
\begin{equation}
\begin{aligned}
    L(\psi;I_x,I_a) &= -E_q[\log p(I_x|v;I_a)] + KL[q_\psi (v|I_x;I_a)||p(v)] \\
    &= \frac{1}{2}\log |2\pi \Sigma_a| + \frac{1}{2} E_q[(I_a - I_x \circ \Phi_v) ^T \Sigma_a^{-1} (I_a - I_x \circ \Phi_v)] + \frac{1}{2}\Big[\log \frac{|\Lambda_v^{-1}|}{|\Sigma_{v|I_x;I_a}|} - 2\\ 
    & + tr(\Lambda_v\Sigma_{v|I_x;I_a}) + u_{v|I_x,I_a}^T \Lambda_v u_{v|I_x,I_a}\Big]\\
    &= \frac{1}{2K} \sum_k (I_a - I_x \circ \Phi_{v_k}) ^T \Sigma_a^{-1} (I_a - I_x \circ \Phi_{v_k}) + \frac{1}{2}\Big[tr(\lambda D \Sigma_{v|I_x;I_a}-\log \Sigma_{v|I_x;I_a})  \\
    & + u_{v|I_x,I_a}^T \Lambda_v u_{v|I_x,I_a} \Big] + const,
\end{aligned}
\end{equation}

\noindent where $K$ is the number of samples $v_k \sim q$ used to approximate the expectation. We use $K=1$. We treat the fixed atlas $I_a$ and the warped individual image $I_x \circ \Phi_v$ as $MN\times 1$ vectors. We denote this naive application of the registration of 2D projected images as the {\it 2D VoxelMorph} method and it serves as a benchmark in our experiments. 

\begin{figure}[t]
    \centering
    \includegraphics[width=0.9\textwidth]{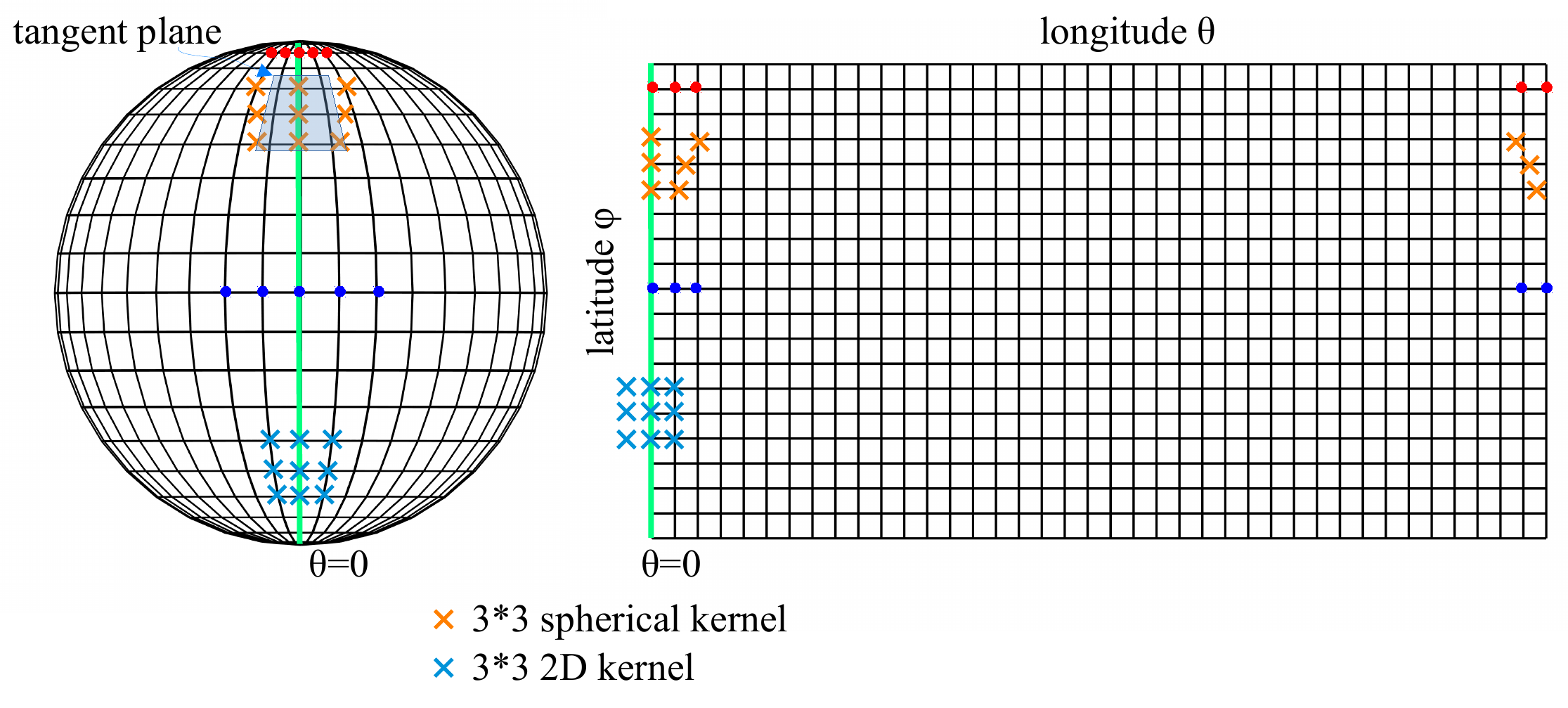}
    \caption{Illustration of a spherical mesh and its corresponding 2D rectangular grids by planar projection. $\varphi$ and $\theta$ denote the longitude and latitude respectively. Regions with higher latitude, e.g. red dots here, have denser samplings compared to regions with lower latitude (e.g. blue dots), yielding greater distortions. The regions crossing $\theta=0$ (green line) become non-adjacent due to the cutting and flattening operations.}
    \label{fig:illustrate}
\end{figure}

Unfortunately, the 2D projection step introduces two main problems, as shown in Fig.~\ref{fig:illustrate}: 
\let\labelitemi\labelitemii
\vspace{-0.5cm}
 \begin{itemize}
       \item[$\bullet$]  varying level of distortions with different latitudes (distortion increases from the equator to the poles); and 
      \item[$\bullet$] inability to represent the periodic property of $\theta$ and the geometry of the poles (an enclosed spherical surface is projected onto a rectangular image region, introducing discontinuities at the image borders).
\end{itemize}

Hence, the {\it 2D VoxelMorph} method over-weights the alignment for near-pole regions, yielding misalignment in most regions even compared to global rigid registration as shown in Section~\ref{sec:results}.

\subsection{Proposed Method: SphereMorph} 

To address the above issues, we propose SphereMorph. We start by defining the registration problem in the spherical domain. The spherical representation of an individual's surface is first rotated for a rigid alignment with the atlas, as in FreeSurfer. The spherical surface is parameterized by the longitude $\theta$ and latitude $\varphi$ and sampled to an $M\times N$ two-dimensional image with a geometric or functional feature, e.g. convexity, assigned as a pixel intensity measure.




\begin{figure}[t]
    \centering
    \includegraphics[width=\textwidth]{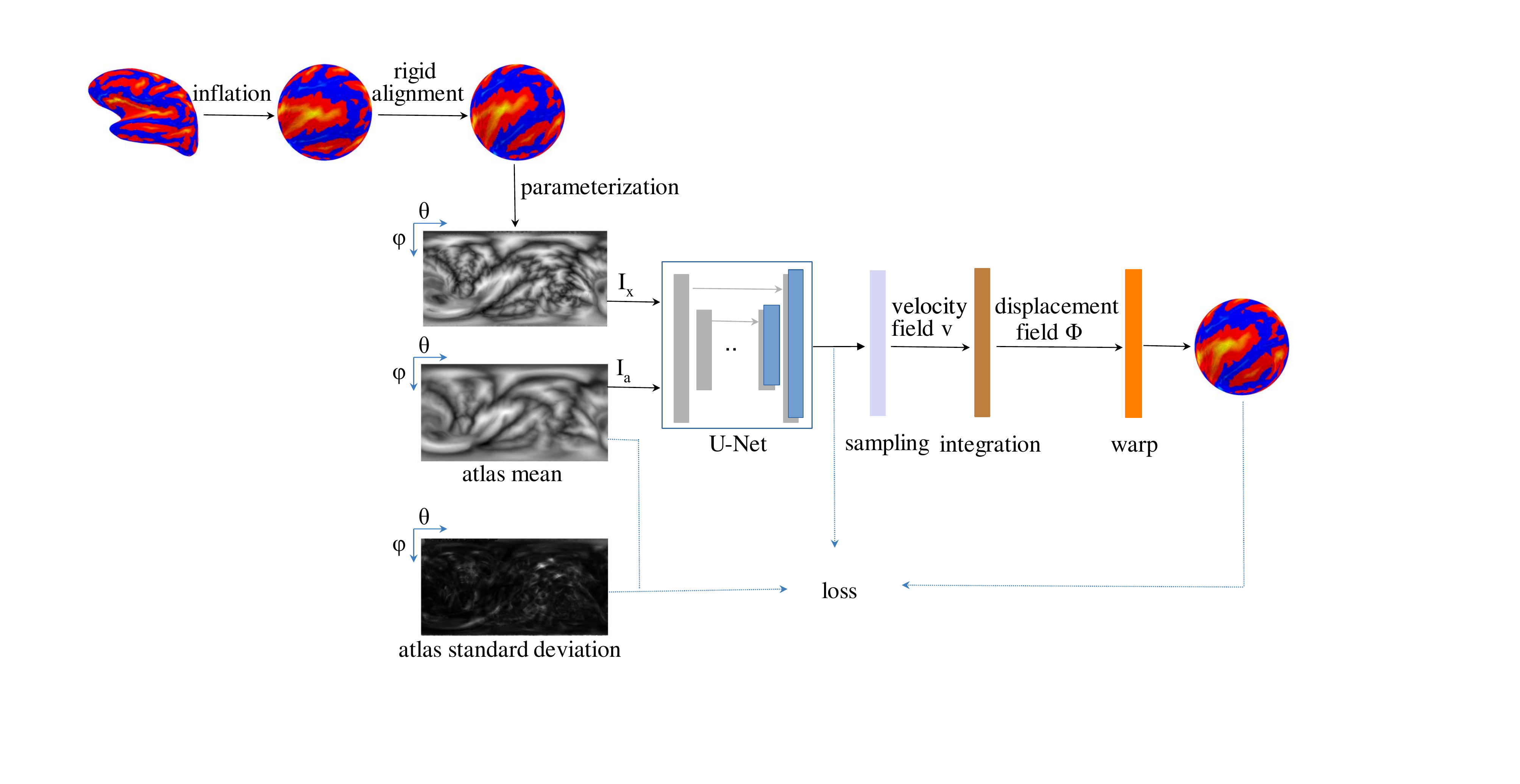}
    \caption{The proposed cortical surface registration workflow (SphereMorph). The input spherical representation is first rigidly aligned to the atlas using convexity patterns and then projected onto a polar coordinate system. The Spherical U-Net core takes parameterized input image $I_x$ and atlas mean image $I_a$ as inputs and estimates the distribution, $\mu_{v|I_a;I_x}$ and $\Sigma_{v|I_a;I_x}$, of the velocity field $v$, which is then sampled and integrated using scaling and squaring steps to generate the deformation field $\Phi_v$.}
    \label{fig:workflow}
\end{figure}
%


\noindent\textbf{Prior correction.} We assume that the displacement field is smooth on the sphere considering the anatomical continuity via a graph Laplacian regularizer. 
We define a neighbour connectivity graph $G_S$ on the spherical manifold and represent the velocity with respect to Cartesian coordinates as a signal defined on this graph. Let $\mathcal{T}$ denote the conversion from polar to Cartesian coordinates, that is $[x,y,z]^T = \mathcal{T}([\theta,\varphi]) = [\sin \varphi \cos \theta, \sin \varphi \sin \theta, \cos \varphi]^T$, then the geodesic velocity at each vertex $Ver_i (\theta_i, \varphi_i), i \in [1,2,..,MN]$ is given by $v' = f(v) = [v_x(\theta_i,\varphi_i), v_y(\theta_i,\varphi_i), v_z(\theta_i,\varphi_i)]^T =  \mathcal{T}(\Phi_v([\theta_i,\varphi_i])) - \mathcal{T}([\theta_i,\varphi_i])$.

Each vertex on the projected image is considered as a node and each grid edge connecting two adjacent nodes as their edge in $G_S$. We connect leftmost and rightmost nodes due to the periodicity in longitude. The weight of the connection between vertices in $G_S$ varies with location to account for the horizontal edge distance on the spherical surface which is proportional to $\sin \varphi$. Thus, we define the weight of each grid edge connecting vertices $Ver_i(\theta_i,\varphi_i)$, $Ver_j(\theta_j, \varphi_j)$ as:
\begin{equation}
    w_{ij} = \left\{
    \begin{array}{lcr}
         1/\sin\varphi, &  \text{if}  &    \varphi_i=\varphi_j, \theta_i \ne \theta_j \\
         1, &  \text{if}   &   \varphi_i \ne \varphi_j, \theta_i=\theta_j.
    \end{array}
    \right.
\end{equation}

We construct the corresponding neighborhood adjacency matrix $A_S$ with entries $A_{S_{ij}} = w_{ij}$ and the degree matrix $D_S$ with diagonal entries $D_{S_{ii}} = \sum_j w_{ij}$. Finally, we denote the Laplacian of this weighted graph as $L_S = D_S - A_S$ and define the covariance of geodesic velocity as $\Sigma_{v'}^{-1} = \Lambda_{S_v} = \lambda L_S$. Intuitively, this formulation can be seen as increasing the regularization near the poles, where the Euclidean distance between mesh nodes is small, and decreasing the weighting near the equator where the Euclidean distances are larger.\\

\noindent\textbf{Distortion correction.} For VoxelMorph, which deals with Euclidean image registration, the sampling of grid points is equally distributed. However, a spherical parameterization leads to denser sampling grids for regions at higher latitudes as shown in Fig.~\ref{fig:illustrate}. Thus, we assign mesh locations from these regions lower weights in computing the data-fitting term, by introducing a diagonal matrix $S \in R^{MN \times MN}$ with each diagonal entry encoding the resampling weight $S_{ii} = \sin \varphi_i$ for each vertex $Ver_i(\theta_i, \varphi_i)$ and model $p(I_x|v;I_a) = \mathcal N(I_x\circ \Phi_v; I_a, S^{-1}\Sigma_a )$. 
The first data-fitting term in Eq.~\eqref{eq:loss1} is then modified as:
\begin{equation}
    -E_q[\log p(I_x|v;I_a)] =  \frac{1}{2K} \sum_k (I_a - I_x \circ \Phi_{v_k}) ^T  \Sigma_a^{-1} S (I_a - I_x \circ \Phi_{v_k}).
    \label{eq:dataloss}
\end{equation}
%

%
%


%
%
%

%
\noindent\textbf{Loss function.} Starting with Eq.~\eqref{eq:loss1} and taking into account the spherical geometry, we arrive at the below objective function:
\begin{equation}
    \begin{aligned}
    L(\psi;I_x,I_a) &= -E_q[\log p(I_x|v;I_a)] + KL[q_\psi (v|I_x;I_a)||p(v)] \\
    &= -E_q[\log p(I_x|v;I_a)] + KL[q_\psi (v'|I_x;I_a)||p(v')] \\
    &= \frac{1}{2K} \sum_k (I_a - I_x \circ \Phi_{v_k}) ^T \Sigma_a^{-1} S (I_a - I_x \circ \Phi_{v_k}) + \frac{1}{2}\Big[tr(\lambda D_S \Sigma_{v'|I_x;I_a}-\log \Sigma_{v'|I_x;I_a})  \\
    & + u_{v'|I_x,I_a}^T \Lambda_{S_v} u_{v'|I_x,I_a} \Big] + const. \\
    \end{aligned}
    \label{eq:ls}
\end{equation}

\noindent where $\mu_{v'|I_x;I_a} = f(\mu_{v|I_x;I_a})$ and $\Sigma_{v'|I_x;I_a} = f(\Sigma_{v|I_x;I_a})$. The first term in Eq.~\eqref{eq:ls} is the data-fitting term, which encourages matching surfaces after warping and the second term drives the posterior to approximate the smoothness prior defined on a spherical grid. 

%
%

\begin{figure}[t]
    \centering
    \includegraphics[width=\textwidth, trim=0 4.5cm 0 5cm,clip]{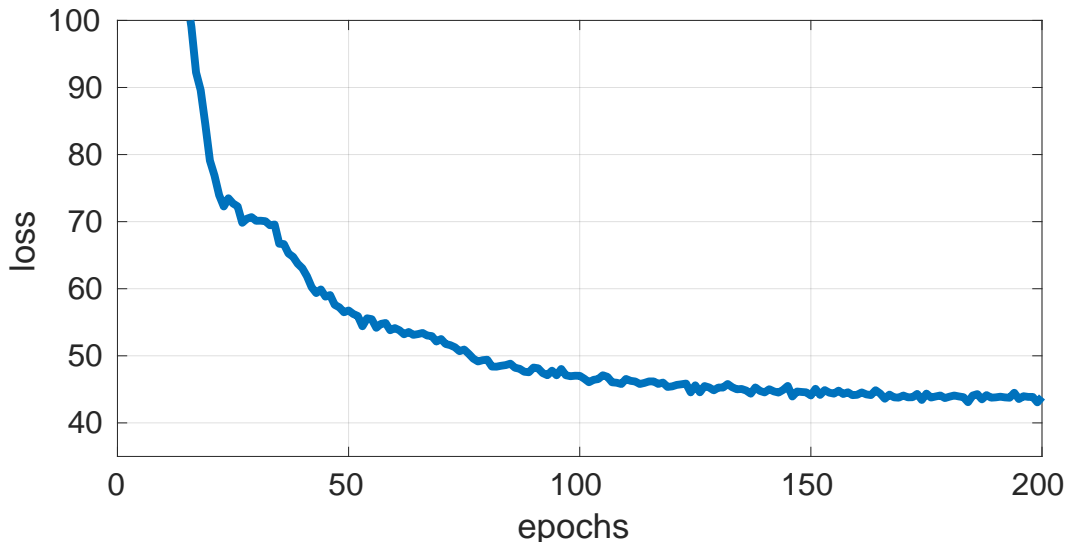}
    \caption{Line plot of the converging loss function over training epochs.}
    \vspace{-0.5cm}
    \label{fig:convergence}
\end{figure}

\noindent\textbf{Network structure.} Figure~\ref{fig:workflow} illustrates the individual stages task fmriof our pipeline. For a given vertex at location $(\theta, \phi)$, we utilize inverse gnomonic projection, which maps points on the tangent plane to the spherical surface as in SphereNet~\cite{coors2018spherenet}, to obtain the corresponding locations on the projected image for the neighbor vertex on its tangent plane. We implement the convolution and pooling operations in each $3\times3$ local tangent patch shown in Figure~\ref{fig:illustrate} and build a UNet core~\cite{ronneberger2015u}, which contains four downsampling and four upsampling layers. Following the sampling layer, seven scaling and squaring operators take the layer output, or velocity field, and return a diffeomorphism $\Phi$. \textit{2D VoxelMorph} uses a dense spatial transformer layer on $(\theta', \phi')$ after displacement to retrieve the warped image while SphereMorph warps the image by computing the interpolation grids as $((\theta'+2\pi)\bmod{2\pi}, \phi')$ for transformer layer. The model is implemented in Keras with a Tensorflow backend and the ADAM optimizer as part of the VoxelMorph package. We set the hyperparameter $\lambda=3\times10^7$, the learning rate to $1\times10^{-5}$, and trained our dense model for $200$ epochs, by which point the loss had converged as shown in Figure~\ref{fig:convergence}. All experiments were conducted on the same workstation with Intel Xeon X5550@2.67GHz and used NVIDIA Tesla P40C for all CNN-based methods. The source code of SphereMorph will be released to the public upon acceptance of this manuscript at \url{
https://github.com/voxelmorph/spheremorph}.


%

\section{Experimental setup}
\label{sec:results}

%


To demonstrate the accuracy and efficiency of the proposed registration framework, SphereMorph, we used two sets of experiments, cortical parcellation and fMRI group analysis, on two independent test data sets. 

\subsection{Data}

\subsubsection{Training data set:}
The spherical atlas from FreeSurfer served as the fixed image in our model. As training data, we used the surface convexity maps of the left hemispheres of 800 randomly selected subjects from the ADNI1 cohort that was released as part of the publicly available Alzheimer's Disease Neuroimaging Initiative (ADNI)~\cite{mueller2005adni} (\url{http://adni.loni.usc.edu/}) and was processed by FreeSurfer v5.1. 
The ADNI was launched in 2003 as a public-private partnership, led by Principal Investigator Michael W. Weiner, MD. The primary goal of ADNI has been to test whether serial magnetic resonance imaging (MRI), positron emission tomography (PET), other biological markers, and clinical and neuropsychological assessment can be combined to measure the progression of mild cognitive impairment (MCI) and early Alzheimer’s disease (AD). The ADNI dataset consists of longitudinal T1-weighted scans from 836 subjects that are divided into four classes: elderly controls (n = 252), early mild cognitive impairment (eMCI, n =215), late MCI (lMCI, n = 176), and AD (n = 193). The subjects were scanned on average 4.8 times (minimum: a single time; maximum: 11 times; 4013 scans in total), with a mean interval between scans equal to 286 days (minimum: 23 days, maximum: 1567 days). The mean age at baseline of the subjects was $75.1\pm6.6$ years. Since the ADNI project spans multiple sites, different scanners were used to acquire the images; further details on the acquisitions can be found at  \url{http://adni.loni.usc.edu/data-samples/adni-data-inventory/}.

\subsubsection{Test data sets:}

(1) We used FreeSurfer-processed MRI scans of 39 subjects from a cohort recruited by the Washington University Alzheimer's Disease Research Center (ADRC)~\cite{van2001buckner}. The MRI scans were acquired on a 1.5T Vision system (Siemens, Erlangen Germany). T1-weighted magnetization-prepared rapid gradient echo (MP-RAGE) scans were obtained according to the following protocol: two sagittal acquisitions, FOV = 224, Matrix = $256\times256$, Resolution = $1\times1\times1.25 mm^3$, TR = 9.7 ms, TE = 4 ms, Flip angle = 10, TI = 20 ms, TD = 200 ms. Two acquisitions were averaged together to increase the contrast-to-noise ratio. For the cortical parcellation experiment, we separated the data set into two: 9 validation subjects and 30 held-out test subjects. All subjects have 34 cortical areas manually annotated~\cite{desikan2006automated}, making them ideal for evaluating registration accuracy.

(2) Additionally, we used another set of 100 unrelated young and healthy subjects from the Human Connectome Project (HCP)~\cite{van2013hcp} as a second test set. The HCP project used state-of-the-art fMRI hardware and acquisition parameters in a sample of highly educated, healthy subjects. For each subject, seven task fMRI sessions were collected, including working memory, gambling, motor, language, social cognition, relational processing and emotional processing, totaling 48:30 $min$ of fMRI data. The acquisition parameters and minimal preprocessing of these data have been described extensively elsewhere~\cite{glasser2013minimal, barch2013function}.


\subsection{Baselines}

We compared our proposed registration method to rigid registration on the sphere (i.e. two rotations) and four other nonlinear registration methods: a 2D version of VoxelMorph, Multimodal Surface Matching (MSM)~\cite{robinson2014msm}, Spherical Demons (SD)~\cite{yeo2010sphericaldemons}, and FreeSurfer (FS) spherical registration~\cite{fischl1999fsreg}. We chose the sulcal depth as the input feature for all registration methods. 
Additionally, we also explored the usage of the curvature and T1/T2 maps together with sulcal depth in the HCP functional group analysis experiments. Additionally, we also explored the usage of the curvature and T1/T2 maps together with sulcal depth in the HCP functional group analysis experiments. We trained 2D VoxelMorph using the same training set as described above and selected the hyperparameter $\lambda$ that yielded the best performance on the validation set. MSM is a surface-based registration approach that offers significant flexibility with regards to the set of features that are used to drive the spatial alignment. MSM drives the deformation via aligning local patches around control points in a multi-resolution fashion. It is implemented on CPU using a fast, multi-resolution, discrete optimisation scheme, offering significant computational speed-up compared to other classical methods. We ran MSM over three resolution levels with five iterations per level. Specifically, we set the regularization parameters as (0.1, 0.2, 0.3) for the parcellation experiments. For the functional group analysis experiments, we relied on the publicly released ‘MSMSulc’ results which were generated by running MSM with high regularization parameters (10, 7.5, 7.5) and using the sulcal depth feature. Compared to the low regularization, MSM with high regularization has been demonstrated to yield lower group alignment in folding patterns but smaller area distortions~\cite{robinson2014msm}, which leads to better functional alignment after following registration steps using “myelin” maps. Spherical Demons, a fast diffeomorphic landmark-free surface registration tool implements the regularization for its objective function via iterative Gaussian smoothing. We explored a range of smoothing iteration numbers (5, 10, 15, 20) to optimize performance and used the results of 10 iterations.


\subsection{Evaluation}

To evaluate registration accuracy, we relied on the resulting spatial transformations to project the atlas parcellation back to individual scan space. For an accurate registration solution, the test subject’s cortical parcellations will resemble the manually outlined versions. In order to quantify how well they match, we computed the Dice overlap coefficient~\cite{dice1945measures}, the overall Mean Minimum Distance (MMD) as well as the individual MMD measures for each anatomical region. The Dice overlap coefficient, $  Dice(M,A) = 2 \cdot \frac{R_{M \cap A}}{R_M + R_A} $, quantifies the surface area overlap ($R_{M \cap A}$) between manual ($R_M$) and automatic method-generated parcels ($R_A$) and the MMD describes the discrepancy between the parcellation boundaries: $MMD(M,A) = 1/N \cdot \Sigma_i d(m_i, a_i)$ where $d(m_i, a_i)$ denotes the Euclidean distance between a vertex $m_i$ on a manual boundary and its corresponding closest vertex $a_i$ on an automatic method-generated boundary. Additionally, we also tested how consistently the various registration methods aligned anatomical features, such as convexity/sulcal depth and curvature.

In the second set of experiments, we mapped all individuals within the group to the HCP's 2mm standard grayordinates space~\cite{glasser2013minimal}, using the displacement field, then computed group maps of task-evoked activations. To evaluate the quality of the group alignment quantitatively, we computed average correlations between the group-average and the projected individual activation maps across 86 task contrasts derived from the seven fMRI tasks.  




 
\section{Results}
\subsection{Computational Efficiency}
Table~\ref{tab:overview} summarizes registration accuracy and computation time for all methods. All the compared methods take spherical cortical surfaces as input, which are generated by brain surface tessellation including fixing topological defects and inflation to the sphere from volumetric segmentation. For a brain surface with 120k-150k vertices, it takes around 0.9h and 0.25h CPU computation time for surface tessellation and inflation, respectively. The comparison of registration time illustrates that the proposed method ran approximately 20 times faster than the conventional registration method in FreeSurfer. On a CPU, the default FreeSurfer pipeline takes around 13 minutes to complete the spherical registration. The total computation time of our proposed framework is approximately 0.74 minutes, including the initial alignment, deep network deployment, and displacement field mapping. With GPU acceleration, the computation time of our method can be reduced to 0.65 minutes, where the deep network deployment is accomplished within a second. Compared with other registration methods including MSM and Spherical Demons (SD), CNN-based methods provide more than an order of magnitude improvement in execution speed.

\begin{figure}
    \centering
    \includegraphics[width=\textwidth]{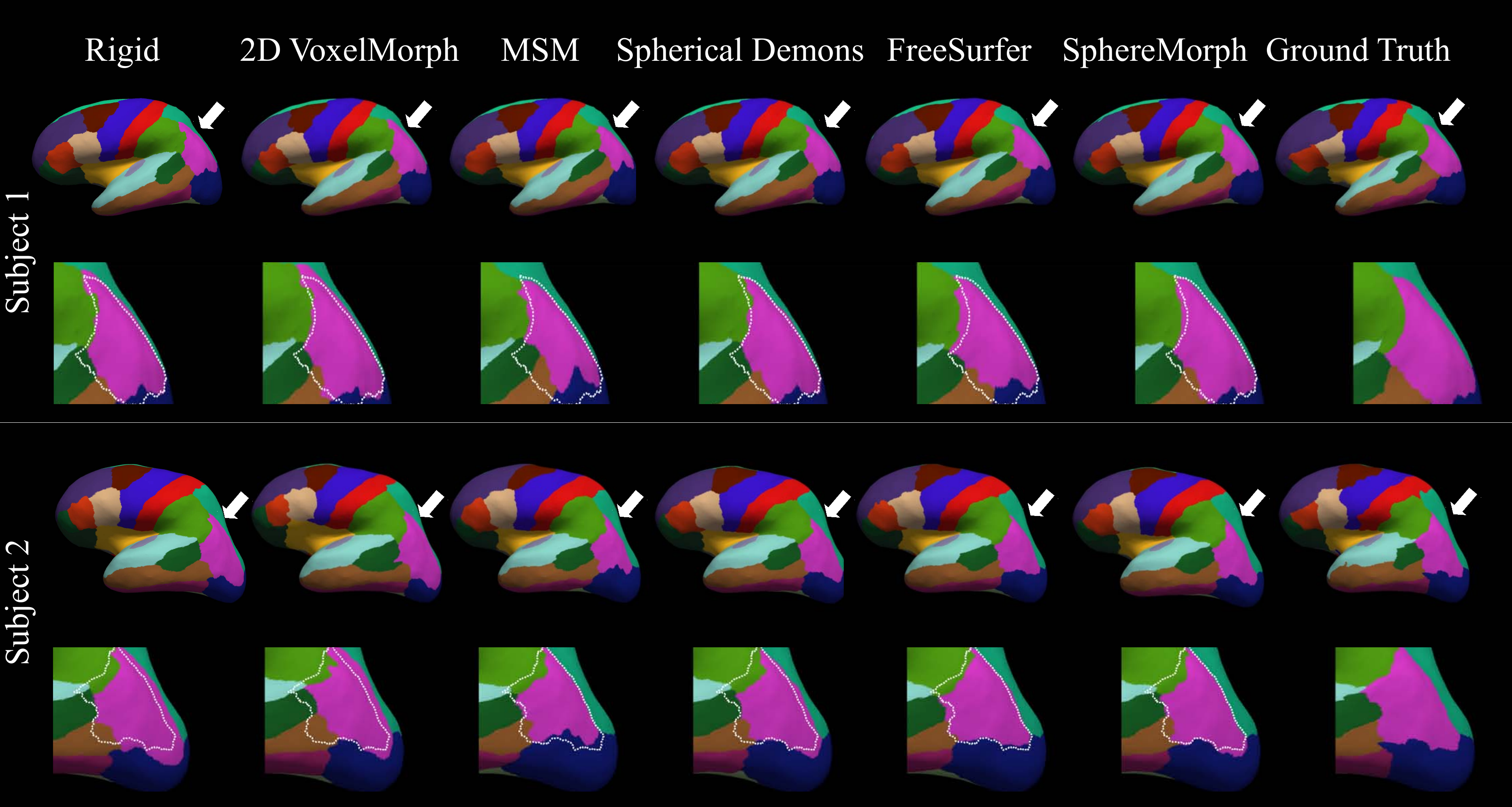}
    \caption{Cortical parcellation results for two example subjects from the ADRC dataset. For each subject, the upper rows show the cortical parcellation estimated by different registration methods and the lower rows give a closer view of the parcellation comparison in the lateral occipital region (white arrows) with manual parcellation boundary superimposed.}
    \label{fig:example}
\end{figure}

\begin{figure}
    \centering
    \includegraphics[width=\textwidth]{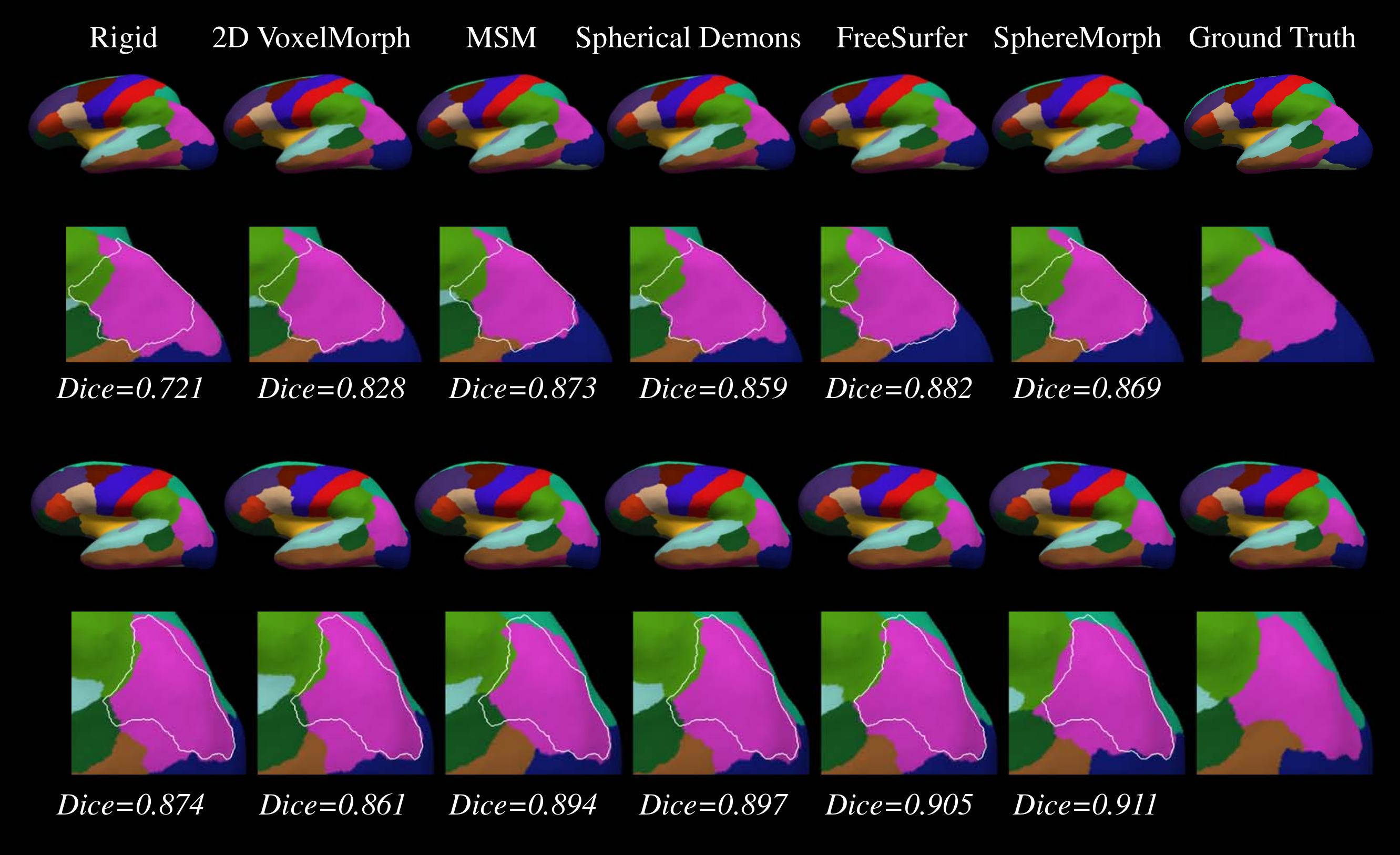}
    \caption{Cortical parcellation results for two subjects where our proposed registration method yields the lowest (top 2 rows) and the highest (bottom 2 rows) overall Dice scores. The overall Dice overlap values for each registration method are displayed below each subfigure. }
    \label{fig:spectrum}
\end{figure}

\begin{figure}[htbp]
    \centering
    \includegraphics[width=\textwidth]{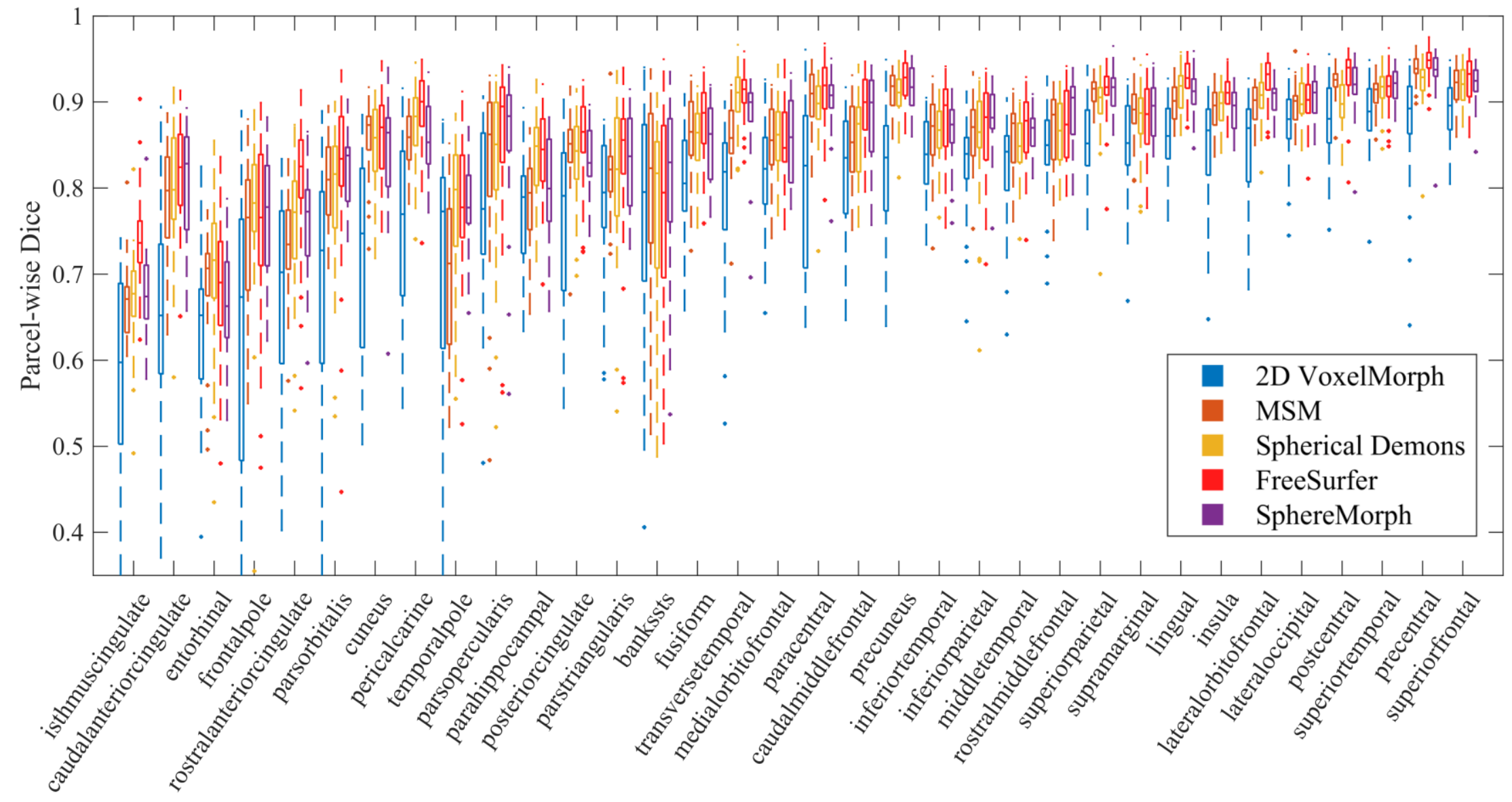}
    \caption{Boxplots of Dice overlap coefficients for all the 34 anatomical structures computed with respect to 2D VoxelMorph, MSM, Spherical Demons, FreeSurfer spherical registration and SphereMorph.}
    \vspace{-0.5cm}
    \label{fig:dice_comp}
\end{figure}

\begin{table}
\caption{Overview of registration accuracy and CPU run time for rigid alignment, 2D VoxelMorph, MSM, Spherical Demons, FreeSurfer spherical registration, and our proposed SphereMorph. SphereMorph performs in a comparable manner with FreeSurfer, while being roughly 20 times faster.}
\label{tab:overview}
\centering
\begin{tabular}{l c c c}
\hline
\hline
Method &  Dice & overall MMD ($mm$) & Time ($min$) \\
\hline
Rigid & $0.840\pm0.029$ & $3.13 \pm 0.43$ & $0.59\pm0.08$\\
2D VoxelMorph &  $0.819 \pm0.027$ & $3.47\pm0.48$& $0.73\pm0.08$  \\
MSM & $0.872\pm0.014$   & $2.63\pm0.27$ & $9.56\pm1.02$  \\
Spherical Demons & $0.881\pm0.009$ &  $2.43\pm0.21$ & $6.15 \pm 0.59$ \\
FreeSurfer &  $0.889\pm0.014$ & $2.24 \pm0.21$ & $13.46\pm3.13$  \\
SphereMorph & $0.882\pm0.014$ & $2.42 \pm0.24$ & $0.74\pm0.09$  \\
\hline
\hline
\end{tabular}
\end{table}

\begin{figure}[t]
    \centering
    \includegraphics[width=\textwidth]{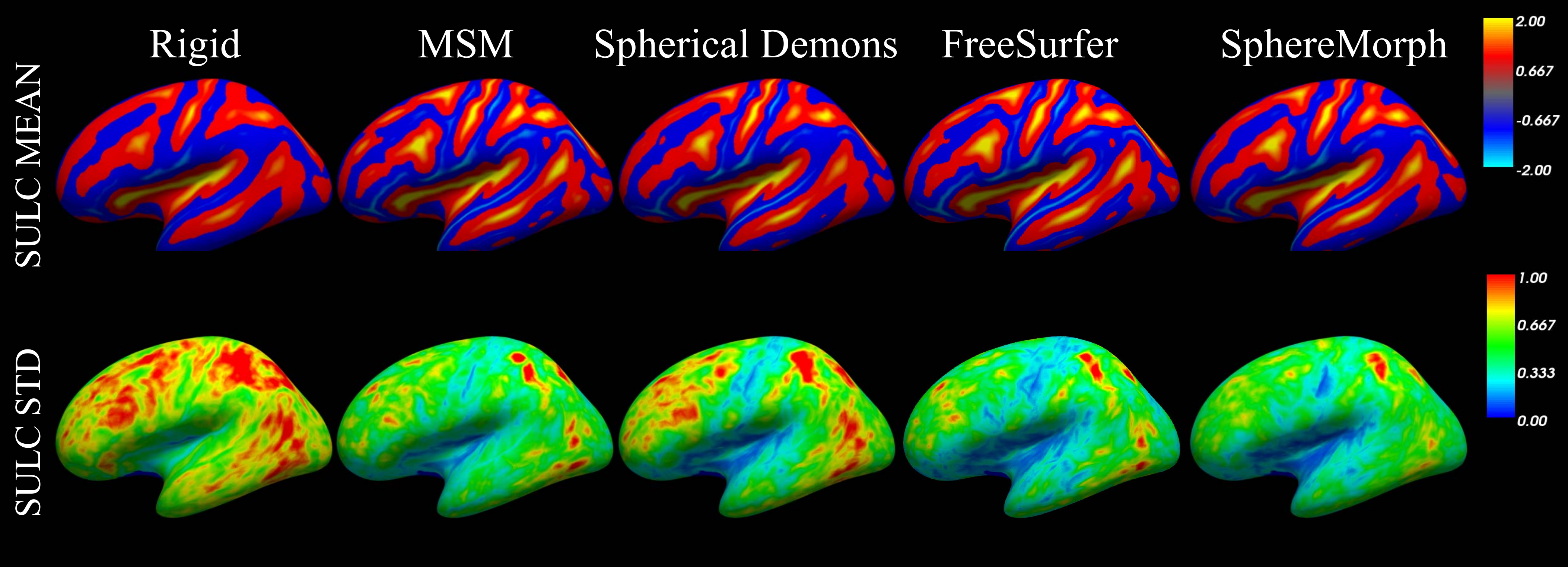}
    \caption{Sulcal depth alignment generated by rigid, MSM, Spherical Demons and SphereMorph. The maps show group mean and standard deviation of convexity from 30 testing subjects.}
    \vspace{-0.5cm}
    \label{fig:sulc1}
\end{figure}

\begin{figure}[b]
    \centering
    \includegraphics[width=\textwidth]{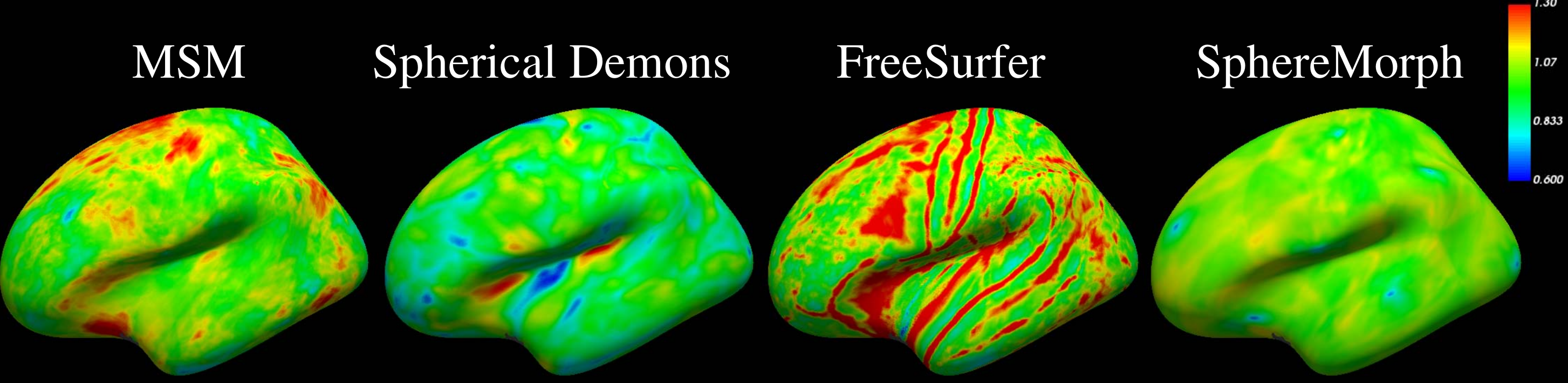}
    \caption{Jacobian maps of registered surfaces, representing areal distortion measurements. Comparing outcomes for all registration solutions, SphereMorph demonstrates the lowest amount of distortion.}
    \label{fig:area_dist}
\end{figure}

\begin{figure}
    \centering
    \includegraphics[width=\textwidth]{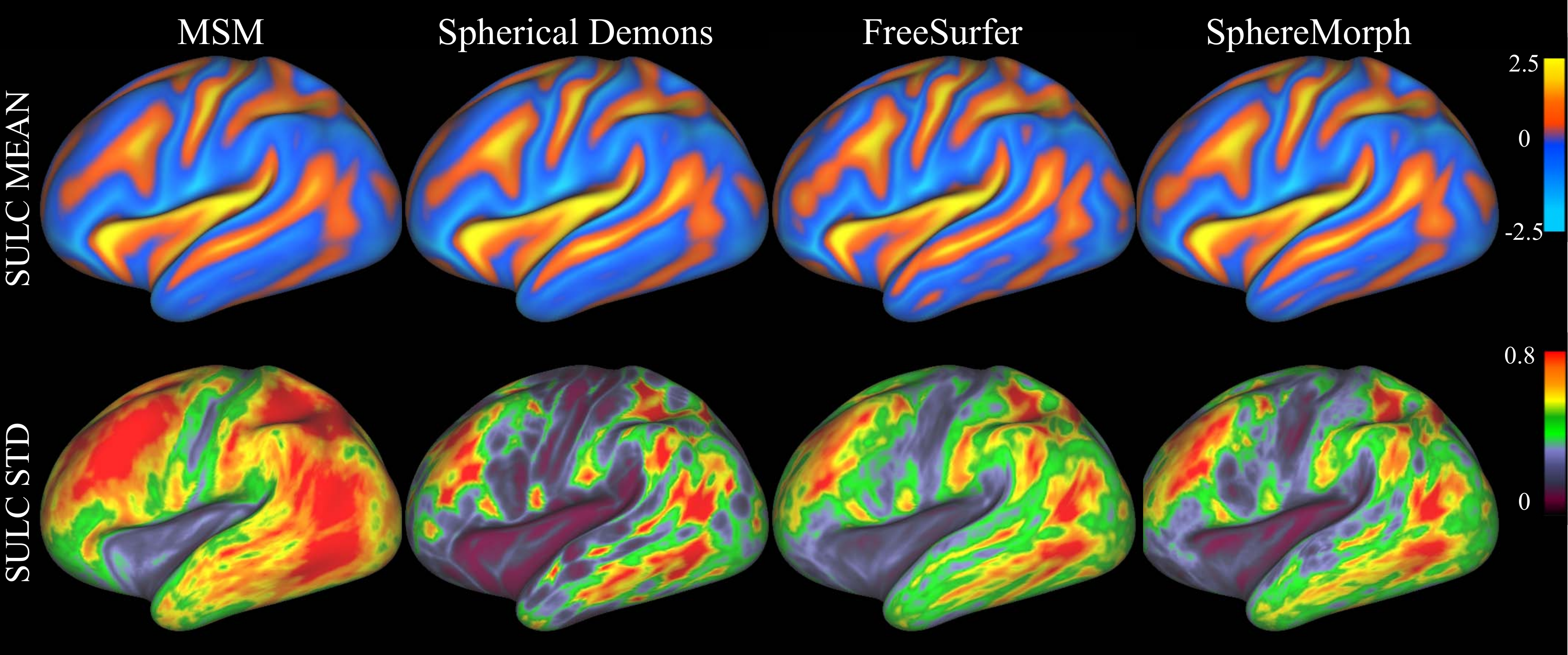} 
    \caption{Spherical Demons, FreeSurfer and SphereMorph for the 100 unrelated subjects in our second test data set. The top panel shows group-average maps and the bottom panel shows group standard deviations.}
    \vspace{-0.5cm}
    \label{fig:sulc2}
\end{figure}

\subsection{Cortical Parcellation experiment}

\subsubsection{Parcellation Accuracy}
Figure~\ref{fig:example} shows representative cortical segmentation results from all the methods on two randomly selected test subjects from the ADRC dataset. The 2D VoxelMorph-estimated annotation exhibits large differences compared to the ground truth in the lateral occipital regions (marked by white arrows), while FreeSurfer and SphereMorph provide results close to the manual annotations. Figure~\ref{fig:spectrum} displays the two subjects for which our proposed method yields the lowest and highest overall Dice values.

Table~\ref{tab:overview} provides an overview of registration accuracy by comparing manual annotations to parcellations generated by the different registration methods, including rigid alignment, 2D Voxelmorph, MSM, SD, FreeSurfer as well as our proposed method. Our proposed method and FreeSurfer achieved the highest accuracy. SphereMorph yields significantly higher overall Dice coefficients than MSM after performing a one-tail Wilcoxon rank sum test on their respective Dice coefficients ($p=0.0318$). The deep learning baseline, 2D VoxelMorph, performed significantly worse. This is due to the fact that this method does not account for the distortions intrinsic to the spherical coordinate system. Figure~\ref{fig:dice_comp} compares parcel-wise Dice overlap coefficient values associated with the different registration methods. Our method produced higher mean Dice overlap coefficients than MSM for all structures except the Entorhinal, Paracentral and Middletemporal, and showed statistically significant improvement in regional Dice overlap coefficient compared to MSM in the Temporalpole ($p=0.0090$), Parahippocampal ($p=0.0102$), Transversetemporal ($p=0.0051$), Caudalmiddlefrontal ($p=0.0237$), Rostralmiddlefrontal ($p=0.0433$) and Lingual ($p=0.0148$) regions.  Compared to Spherical Demons, SphereMorph produced higher mean regional Dice coefficients in 26 out of 34 regions. The regions where this was not the case are the Entorhinal, Rostralanteriorcingulate, Parahippocampal, Fusiform, Transversetemporal, Insula and Lateralorbitofrontal areas. Pairwise t-tests show that SphereMorph compared to SD generated statistically significantly higher Dice scores in the Postcentral ($p=0.0015$) and Rostralmiddlefrontal ($p=0.0191$) areas while SD outperformed SphereMorph in the Transversetemporal ($p=0.0245$) region. In all other areas the performance difference was not statistically significant. Postcentral and Rostralmiddlefrontal regions exhibit larger standard deviations ($0.12\pm0.05$ and $0.15\pm0.09$ respectively) as shown in the atlas while the Transversetemporal region has higher agreement in sulcal depth (atlas standard deviation $0.05\pm0.02$) across subjects. The inverse-variance weighting in SphereMorph makes it more flexible in regions with large group variance, leading to better parcellation results in these regions when compared to SD. SD uses the sum of squared differences as a similarity estimate and yields good parcellation in regions with high agreement in sulcal depth.

\subsubsection{Group average sulcal maps}
To evaluate the performance on a finer scale, we also computed the group mean sulcal depth maps after registration for the test data. The resulting mean and standard deviation maps are displayed in Figure~\ref{fig:sulc1}. We assume that a better group alignment leads to a sharper group mean and smaller group variation. As expected, the group mean maps provide more detailed information for all nonlinear registration methods than rigid alignment. Moreover, all these methods show smaller standard deviations, suggesting that they provide better alignment in convexity. 

All nonlinear registration methods exhibited similar distributions of sulcal variations across brain regions. Specifically, the pre-central, post-central and insula regions show lower standard deviation, suggesting higher agreement of cross-subject convexity in these regions. 


\subsubsection{Robustness Analysis}

We investigated the impact of pole positioning on the registration accuracy for SphereMorph by projecting the sphere using 9 different north pole locations spanning $\theta \in (0, \pi/2, \pi)$, $\phi \in (0, \pi/6, \pi/3, \pi/2)$ and registering the corresponding 2D planar images with corresponding atlas data. 
We generated the deformed sphere and then re-computed all the region- and distance-based metrics. To evaluate the robustness with respect to different projection centers, we conducted analysis of variance (ANOVA) between all the computed evaluation metrics for these 9 groups.
None of the ANOVA analyses (overall Dice: $F=0.2, p=0.9842$, overall MMD: $F=0.93, p=0.4821$) found any significant differences between the 9 groups of cortical parcellations, indicating that the registration accuracy is not sensitive to the arbitrary location of the poles.

\subsubsection{Areal Distortion Analysis}
We calculated the Jacobian map for registered surfaces as an areal distortion measure, where a value close to one indicates a small distortion. Figure~\ref{fig:area_dist} compares the distortion maps for all compared registration methods, suggesting least amount of distortion introduced by SphereMorph. Additionally, we measured the percentage of vertices with a non-positive Jacobian determinant for SphereMorph in the case of a randomly selected subject and it resulted in 0.63\%. While we model the deformation field as a diffeomorphism in the continuous case, discretization and numerical errors can lead to a small fraction of locations with Jacobians that are not positive definite.

\begin{figure}[!ht]
    \centering
    \includegraphics[width=0.6\textwidth]{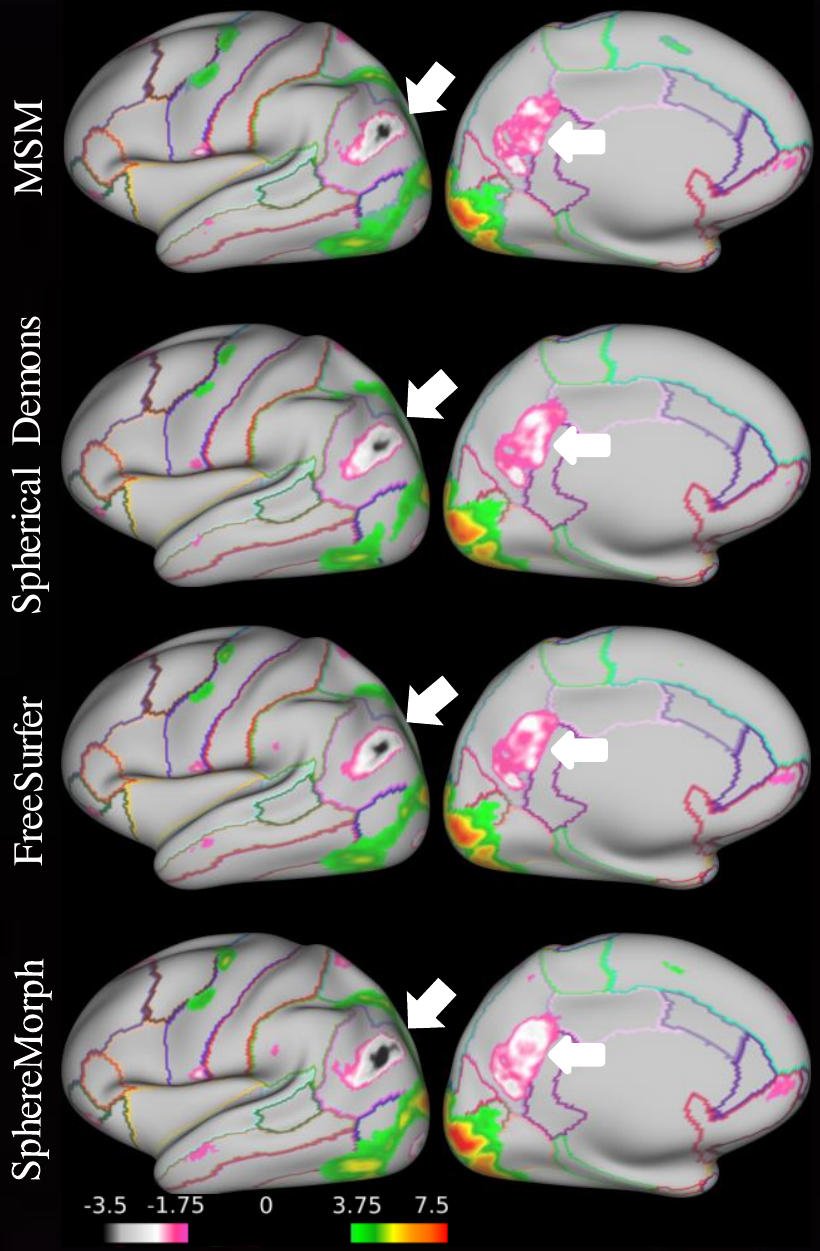}
    \caption{Task fMRI alignment resulting from convexity-driven spatial registration using MSM, Spherical Demons, FreeSurfer and SphereMorph. The maps show group activation results from 100 subjects for the Gambling task reward contrasts (with the outline of Freesurfer-based cortical parcellations for easier interpretation). SphereMorph improves the functional alignment across subjects over MSM, particularly in the inferiorparietal and precuneus regions, indicated by the arrows.}
    \vspace{-0.5cm}
    \label{fig:gamblingtask}
\end{figure}

\subsection{Cross-subject alignment of fMRI activation maps}
\subsubsection{Group average maps}
Figure~\ref{fig:sulc2} illustrates the quality of group-wise alignment of convexity after MSM, Spherical Demons, FreeSurfer, and our method were run on our second test data set. Our method yielded smaller variations within the group than MSM ($p=0.0013$), just as in the case of the cortical parcellation experiments. For these experiments, we used default regularization parameters for FreeSurfer and optimized the regularization level of Spherical Demons via exploring the number of smoothing iterations for best parcellation performance. Regarding MSM, we relied on the HCP-distributed MSM registration results that were driven only by folding patterns with a high regularization, as suggested in~\cite{robinson2014msm}, where such settings lead to better functional alignment when followed by a registration step using “myelin” maps. This choice lead to the high group variance displayed on Figure~\ref{fig:sulc2}. In order to make this comparison more fair, we also ran MSM using a combination of the T1/T2 map with the folding patterns for an evaluation of accuracy in functional alignments as explained below. We denote the MSM registration version incorporating folding patterns along with the T1/T2 map as MSM-T1/T2 and the SphereMorph version as SphereMorph-T1/T2.


While we expect folding patterns to be predictive of functional areas, this relationship is variable and complex. In order to assess how well the various methods align functionally homologous regions across subjects, we computed group average functional activation maps for all 86 tasks using registration driven by sulcal depth information. Figure~\ref{fig:gamblingtask} compares group activation results from the 100 subjects for the Gambling reward contrast. Our method improves the functional alignment across subjects over MSM, particularly in the inferiorparietal and precuneus regions, indicated by the arrows. To evaluate the group alignment performance quantitatively, we computed the average correlations between the group-average and individual activation maps after registration across 86 task contrasts derived from seven tasks. All results in Table~\ref{tab:func_comp} were computed using registration results driven by sulcal depth maps except for MSM-T1/T2 and SphereMorph-T1/T2, where the T1/T2 map was also incorporated in order to drive the registration.

\begin{figure}[htbp]
    \centering
    \makebox[\textwidth][c]{\includegraphics[width=0.8\textwidth]{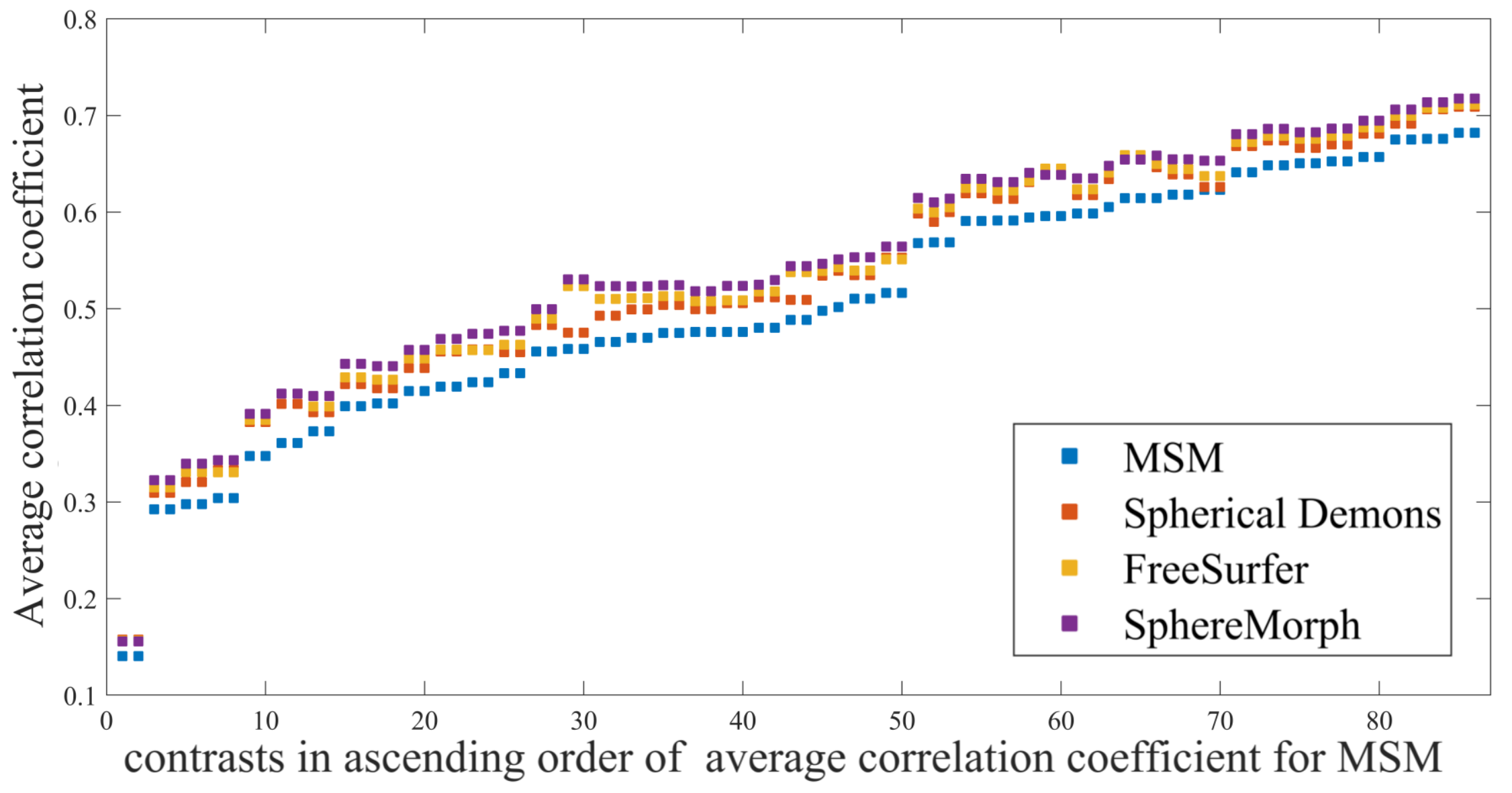}}
    \caption{Comparison in the correlation coefficients computed between the group-average and individual activation maps after registration across 86 task contrasts. Higher coefficients suggest better agreement in task activations across subjects.}
    \vspace{-0.5cm}
    \label{fig:ccor}
\end{figure}

Figure~\ref{fig:ccor} displays average correlations for the different registration methods. For all 86 contrasts, SphereMorph resulted in significantly higher correlation coefficients than MSM (increase of $0.043 \pm 0.009, p=0.028$ and relative increase of $8.99\pm 2.85 \%, p=0.0024$).

\begin{table}
\caption{Comparison of group functional alignment for MSM, Spherical Demons, FreeSurfer spherical registration, and our proposed method SphereMorph. Correlation coefficients are calculated for the average correlation coefficients of all 86 contrasts. MSM-T1/T2 and SphereMorph-T1/T2 denote the incorporation of the T1/T2 map along with the sulcal depth feature. All other methods only rely on the sulcal depth.}
\vspace{5pt}
\label{tab:func_comp}
\centering
\begin{tabular}{l c c c}
\hline
\hline
Method &  correlation coefficients & difference with MSM & relative difference ($\%$)\\
\hline
MSM               & $0.5037 \pm 0.1256$ & - & - \\
Spherical Demons  & $0.5298 \pm 0.1258$ & $0.0261 \pm 0.0082$ & $5.60  \pm 2.35$\\
FreeSurfer        & $0.5372 \pm 0.1259$ & $0.0335 \pm 0.0093$ & $7.12  \pm 2.62$\\
SphereMorph       & $0.5460 \pm 0.1254$ & $0.0424 \pm 0.0090$ & $8.99  \pm 2.85$\\
MSM-T1/T2         & $0.5456 \pm 0.1310$ & $0.0419 \pm 0.0081$ & $8.65  \pm 1.89$\\
SphereMorph-T1/T2 & $0.5535 \pm 0.1243$ & $0.0498 \pm 0.0164$ & $10.64 \pm 4.46$\\ 
\hline
\hline
\end{tabular}
\end{table}

\subsubsection{Alignment performance analysis using multi-modality features}
\begin{figure}[htbp]
    \centering
    \makebox[\textwidth][c]{\includegraphics[width=1\textwidth]{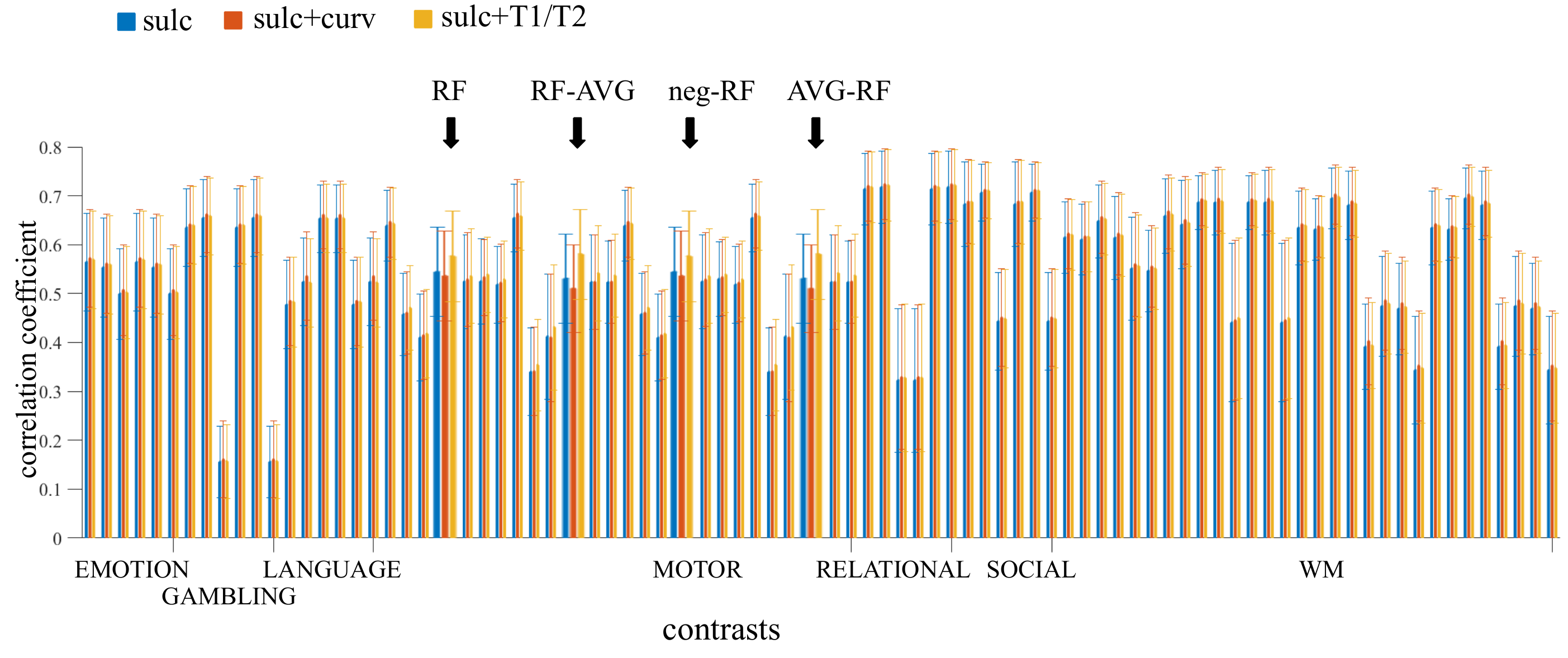}}
    \caption{Comparison of agreement in 86 task activations using different features. We take the `sulc' and `curv' atlases from FreeSurfer and compute a `T1/T2' atlas from three rounds of CNN registration within group. Contrasts showing significant improvement ($p < 0.01$) after incorporation of the T1/T2 feature are pointed out by arrows. }
    \vspace{-0.5cm}
    \label{fig:diffinputs}
\end{figure}

\begin{figure}[htbp]
    \centering
    \makebox[\textwidth][c]{\includegraphics[width=0.65\textwidth]{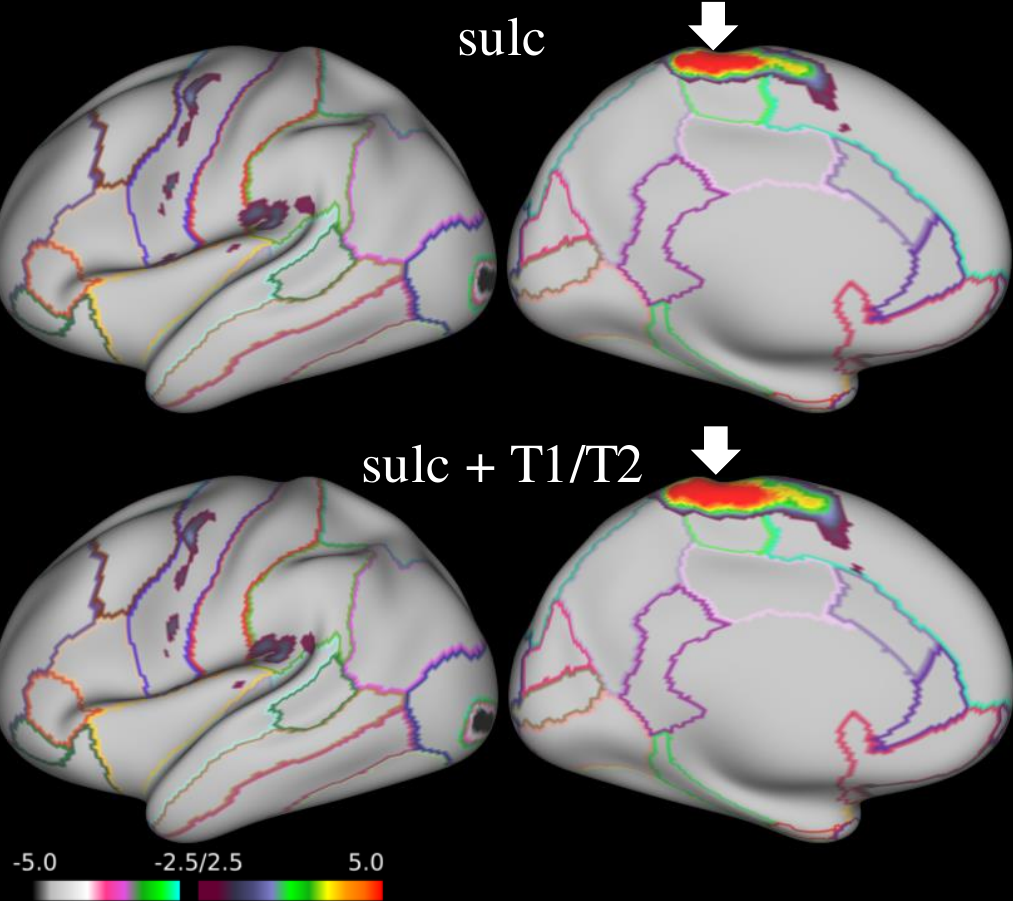}} 
    \caption{Task fMRI alignment driven by using different feature maps (sulcal depth and sulcal depth + T1/T2). The maps show group activation results from 100 subjects for the MOTOR task: RF contrasts, with the outline of Freesurfer-based cortical parcellations for easier interpretation. The arrows highlight the paracentral and superiorfrontal areas, where using the T1/T2 map leads to a larger region with positive response compared to using sulcal depth as the only input.}
    \vspace{-0.5cm}
    \label{fig:rftask}
\end{figure}

We evaluated the performance for the proposed SphereMorph using curvature and (separately) T1/T2 features to their respective atlases, using our sulcal depth alignment as initialization. We computed a `T1/T2' atlas from three rounds of CNN registration within group. Figure~\ref{fig:diffinputs} compares the group agreements in task activation for only sulc and two cascade processes using respective curvature and T1/T2 maps. Using T1/T2 as input post sulcal depth-based registration significantly improves the group agreement level ($p<0.01$) in four MOTOR tasks, including RF, RF-AVG, neg-RF and AVG-RF, as indicated by the arrows. All four tasks are related to right finger tapping. Figure~\ref{fig:rftask} displays the group average activation for RF contrast. Using the T1/T2 map leads to a larger region with
positive response in the paracentral and superiorfrontal areas (marked by arrows) compared to using sulcal depth as the only input.

\section{Discussions and Conclusions}
\label{sec::diss}

In this paper we present a learning-based method, SphereMorph, for registering cortical surfaces and investigate its performance using two sets of experiments, by comparing it to rigid alignment, 2D VoxelMorph, MSM, Spherical Demons, and FreeSurfer. The proposed SphereMorph yields results that are comparable or superior to the state-of-the-art methods for alignment of folding patterns, cortical parcellation and functional alignment, while offering approximately a $20\times$ computational speedup, showing the accuracy and efficiency of our method for cortical registration. Different from existing spherical networks that learn a node-wise (e.g. cortex parcellation) or subject-wise measure (e.g. gender or abnormality classification), our registration framework is more flexible and can be applied for both parcellation as well as various group analysis tasks.

Compared to conventional registration methods, SphereMorph takes 2D parameterized images as input, outputs the deformation field in a 2D canonical space, then warps the sphere in the original Cartesian space. We directly address two issues associated with the 2D projection: the effects of substantial distortions introduced by the parameterization and the violation of continuity at the borders of the 2D plane (i.e. the imposition of spherical topology). To account for the distortion introduced by the parameterization, we modify the data likelihood term and construct the deformation velocity on a graph weighted by the metric tensor of the parameterization. We use the same weights as in existing work by Khasanova and Frossard~\cite{khasanova2017weight} to construct our graph as it has been shown to be capable of encoding the geometry of an omnidirectional camera in the final feature representation of an image. In the implementation of network structure, we encode the neighborhood information by leveraging a $3\times3$ spherical kernel defined in SphereNet~\cite{coors2018spherenet} instead of a conventional 2D kernel for the network convolution and pooling operations. Compared to the spherical kernel defined in~\citep{su2017learning}, the SphereNet kernel is able to handle both the discontinuity problem and planar distortions. While other existing spherical kernels~\cite{seong2018geometric,jiang2019sphericalunet,zhao2019sphereunet} defined on the 3D mesh could be used for the spherical operations in our networks, given that we model the registration problem based on the longitude-latitude representation, an additional interpolation step would need to be introduced for them, potentially introducing error.

We then use a scaling-and-squaring layer to obtain the exponential map in the spherical domain and this can be efficiently implemented on the GPU, without the density of the nodes significantly impacting runtime~\cite{dalca2019varreg}. In addition, we modify the spatial transformer layer to represent the periodicity of longitude in spherical coordinates.  
Combining these adaptations, our model shows a higher agreement with manual parcellations compared to 2D VoxelMorph, which does not account for distortions and topological changes induced by the 2D parameterization. For 2D VoxelMorph, each point on the rectangular grid contributes equally to the registration, resulting in the alignment of regions near the poles affecting the energy functional more than other regions with the same area in the Euclidean embedding space. Thus, the optimized deformation is over-fitted in these regions. Our experimental results confirm this when comparing the automatically generated parcellations to the manual ones. With an excessive weighting of the alignment of these polar regions, 2D VoxelMorph performs even worse than the initial alignment.


Our proposed registration framework is applicable to any signal or feature that can be represented or sampled onto the cortical surface. However, in the cortical parcellation experiment, the ADRC dataset only has structural scans. Hence, we trained our model 
as well as MSM and SD using convexity values to drive the registrations. FreeSurfer originally trained its parcellation atlas using the manual annotations on the ADRC data set and yields the best parcellation performance but longest execution time among all the compared methods. Compared to FreeSurfer, which uses a line minimization optimization strategy to obtain a dense displacement field, Spherical Demons reduces the registration problem to a Gauss-Newton optimization step of a non-linear least-squares problem and a displacement field smoothing operation by simple convolution, resulting in faster computation time. We found no significant difference between our proposed SphereMorph and FreeSurfer while SphereMorph shows significant smaller group variations and higher overall Dice coefficients when compared to MSM, suggesting SphereMorph achieves better structural alignment than MSM.


In addition to the accuracy of the structural alignment, when using the same feature (i.e. convexity) for registration, both qualitative and quantitative evaluation results for the task-related experiments demonstrate that the proposed method also yields higher agreement of functional regions across subjects than the current registration method in the Human Connectome Project (HCP) pipeline~\cite{glasser2013minimal}. SphereMorph improves the within-group correlation coefficients significantly for all tasks compared with MSM. The group agreement in convexity is also higher after SphereMorph registration than with MSM. This implies the relationship between cortical folding patterns and boundaries of functional areas as demonstrated in previous work~\cite{hinds2008accurate}. However, using folding-based features alone may not be sufficient to provide an accurate functional alignment across the entire cortex due to regions with highly variable folding patterns across subjects, as well as structure-function variability. Thus, in the functional alignment experiments, we take the sulcal depth registration as initialization and then use cortical T1/T2 maps as input of our network for a multi-step alignment. The cortical T1/T2 maps are computed based on the ratio of T1-weighted to T2-weighted images which correlates with many functionally distinct areas in individual subjects~\cite{glasser2011mapping}. The combination of T1/T2 and sulcal depth shows significant improvement in group alignment of MOTOR-related tasks, indicating the correlation between T1/T2 and MOTOR-related regions. SphereMorph, just like FreeSurfer, uses an inverse variance weighting scheme to compute the data similarity term. At final evaluation and in comparison with other methods not using such a weighting cheme, SphereMorph can be at a disadvantage due to the presence of areas that are not optimized for alignment. Interestingly though, we found that SphereMorph yields better parcellation outcomes than SD in the Postcentral and Rostralmiddlefrontal areas, which are regions of high variance in the atlas, while SD performs better in the Transversetemporal region that is of lower variance. Consistent with these findings, Table~\ref{tab:func_comp} demonstrates higher accuarcy functional alignment for SphereMorph compared to SD, while SD yields smaller group variations in folding patterns. This implies that increased structural variability is a marker of reduced structure-function coupling. Thus we believe that using inverse-variance weighting helps improving the functional alignment.

All experiments in this paper focus on subject-to-atlas registration. We will explore the performance of our network for inter-subject registration in the future. In addition, our proposed model can include other contrasts in addition to the convexity metrics used above to drive the registration. Existing studies have shown the improvement in group registration accuracy relying on features generated from resting-state functional MRI (rfMRI) after conventional convexity-driven registration ~\cite{robinson2014msm,tong2017functional}. In the future, we will investigate the use of these to further improve the accuracy of our technique. 

\section{Acknowledgement}
\justifying
\begin{sloppypar}
Support for this research was provided in part by the BRAIN
Initiative Cell Census Network grant U01MH117023, the Eunice Kennedy Shriver National Institute of Child Health and Human Development (NICHD) (5R21HD95338-02, 5R01HD085813-04, 5R01 HD065762-09, 5R01HD093578-03, 5R01EB024343-03), the National Institute for Biomedical Imaging and Bioengineering (P41EB015896, 1R01EB023281, R01EB006758, R21EB018907, R01EB019956, 5R03EB022754-02), the National Institute on Aging (1R56AG064027, 5R01AG008122, R01AG016495), the National Institute of Mental Health 5U01MH109589-04,  the National Institute of Diabetes and Digestive and Kidney Diseases (1-R21-DK-108277-01), the National Institute for Neurological Disorders and Stroke (R01NS0525851, R21NS072652, R01NS070963, R01NS083534, 5U01NS086625, 5U24NS10059103, R01NS105820), and was made possible by the resources provided by Shared Instrumentation Grants 1S10RR023401, 1S10RR019307, and 1S10RR023043. Additional support was provided by the NIH Blueprint for Neuroscience Research (5U01-MH093765), part of the multi-institutional Human Connectome Project. In addition, BF has a financial interest in
CorticoMetrics, a company whose medical pursuits focus on brain
imaging and measurement technologies. BF's interests were reviewed and
are managed by Massachusetts General Hospital and Partners HealthCare in accordance with their conflict of interest policies. Data collection and sharing for this project was also partially funded by the Alzheimer's Disease Neuroimaging Initiative (ADNI) (National Institutes of Health Grant U01 AG024904) and DOD ADNI (Department of Defense award number W81XWH-12-2-0012). ADNI is funded by the National Institute on Aging, the National Institute of Biomedical Imaging and Bioengineering, and through generous contributions from the following: AbbVie, Alzheimer’s Association; Alzheimer’s Drug Discovery Foundation; Araclon Biotech; BioClinica, Inc.; Biogen; Bristol-Myers Squibb Company; CereSpir, Inc.; Cogstate; Eisai Inc.; Elan Pharmaceuticals, Inc.; Eli Lilly and Company; EuroImmun; F. Hoffmann-La Roche Ltd and its affiliated company Genentech, Inc.; Fujirebio; GE Healthcare; IXICO Ltd.; Janssen Alzheimer Immunotherapy Research \& Development, LLC.; Johnson \& Johnson Pharmaceutical Research \& Development LLC.; Lumosity; Lundbeck; Merck \& Co., Inc.; Meso Scale Diagnostics, LLC.; NeuroRx Research; Neurotrack Technologies; Novartis Pharmaceuticals Corporation; Pfizer Inc.; Piramal Imaging; Servier; Takeda Pharmaceutical Company; and Transition Therapeutics. The Canadian Institutes of Health Research is providing funds to support ADNI clinical sites in Canada. Private sector contributions are facilitated by the Foundation for the National Institutes of Health (\url{www.fnih.org}). The grantee organization is the Northern California Institute for Research and Education, and the study is coordinated by the Alzheimer’s Therapeutic Research Institute at the University of Southern California. ADNI data are disseminated by the Laboratory for Neuro Imaging at the University of Southern California.
\end{sloppypar}




\bibliographystyle{elsarticle-num}

\bibliography{main}

\begin{thebibliography}{10}
\expandafter\ifx\csname url\endcsname\relax
  \def\url#1{\texttt{#1}}\fi
\expandafter\ifx\csname urlprefix\endcsname\relax\def\urlprefix{URL }\fi
\expandafter\ifx\csname href\endcsname\relax
  \def\href#1#2{#2} \def\path#1{#1}\fi

\bibitem{van2004integration}
N.~Van~Atteveldt, E.~Formisano, R.~Goebel, L.~Blomert, Integration of letters
  and speech sounds in the human brain, Neuron 43~(2) (2004) 271--282.

\bibitem{frost2012measuring}
M.~A. Frost, R.~Goebel, Measuring structural--functional correspondence:
  spatial variability of specialised brain regions after macro-anatomical
  alignment, NeuroImage 59~(2) (2012) 1369--1381.

\bibitem{fischl1999fsreg}
B.~Fischl, M.~I. Sereno, R.~B. Tootell, A.~M. Dale, High-resolution
  intersubject averaging and a coordinate system for the cortical surface,
  Human Brain Mapping 8~(4) (1999) 272--284.

\bibitem{yeo2010sphericaldemons}
B.~T. Yeo, M.~R. Sabuncu, T.~Vercauteren, N.~Ayache, B.~Fischl, P.~Golland,
  Spherical demons: fast diffeomorphic landmark-free surface registration, IEEE
  Transactions on Medical Imaging 29~(3) (2010) 650--668.

\bibitem{conroy2013inter}
B.~R. Conroy, B.~D. Singer, J.~S. Guntupalli, P.~J. Ramadge, J.~V. Haxby,
  Inter-subject alignment of human cortical anatomy using functional
  connectivity, NeuroImage 81 (2013) 400--411.

\bibitem{tardif2015multi}
C.~L. Tardif, A.~Sch{\"a}fer, M.~Waehnert, J.~Dinse, R.~Turner, P.-L. Bazin,
  Multi-contrast multi-scale surface registration for improved alignment of
  cortical areas, NeuroImage 111 (2015) 107--122.

\bibitem{robinson2014msm}
E.~C. Robinson, S.~Jbabdi, M.~F. Glasser, J.~Andersson, G.~C. Burgess, M.~P.
  Harms, S.~M. Smith, D.~C. Van~Essen, M.~Jenkinson, {MSM}: a new flexible
  framework for multimodal surface matching, NeuroImage 100 (2014) 414--426.

\bibitem{sotiras2013deformable}
A.~Sotiras, C.~Davatzikos, N.~Paragios, Deformable medical image registration:
  A survey, IEEE Transactions on Medical Imaging 32~(7) (2013) 1153--1190.

\bibitem{thirion1998image}
J.-P. Thirion, Image matching as a diffusion process: an analogy with maxwell's
  demons, Medical Image Analysis 2~(3) (1998) 243--260.

\bibitem{ashburner2007DARTEL}
J.~Ashburner, A fast diffeomorphic image registration algorithm, NeuroImage
  38~(1) (2007) 95--113.

\bibitem{zhang2017frequency}
M.~Zhang, R.~Liao, A.~V. Dalca, E.~A. Turk, J.~Luo, P.~E. Grant, P.~Golland,
  Frequency diffeomorphisms for efficient image registration, in: IPMI,
  Springer, 2017, pp. 559--570.

\bibitem{krebs2017robust}
J.~Krebs, T.~Mansi, H.~Delingette, L.~Zhang, F.~C. Ghesu, S.~Miao, A.~K. Maier,
  N.~Ayache, R.~Liao, A.~Kamen, Robust non-rigid registration through
  agent-based action learning, in: MICCAI, Springer, 2017, pp. 344--352.

\bibitem{sokooti2017nonrigid}
H.~Sokooti, B.~de~Vos, F.~Berendsen, B.~P. Lelieveldt, I.~I{\v{s}}gum,
  M.~Staring, Nonrigid image registration using multi-scale {3D} convolutional
  neural networks, in: MICCAI, Springer, 2017, pp. 232--239.

\bibitem{yang2016pred}
X.~Yang, R.~Kwitt, M.~Niethammer, Fast predictive image registration, in: Deep
  Learning and Data Labeling for Medical Applications, Springer, 2016, pp.
  48--57.

\bibitem{balakrishnan2019tmi}
G.~Balakrishnan, A.~Zhao, M.~Sabuncu, J.~Guttag, A.~V. Dalca, Voxelmorph: a
  learning framework for deformable medical image registration, IEEE
  Transactions on Medical Imaging 38 (2019) 1788--1800.

\bibitem{dalca2018miccai}
A.~V. Dalca, G.~Balakrishnan, J.~Guttag, M.~R. Sabuncu, Unsupervised learning
  for fast probabilistic diffeomorphic registration, MICCAI 11070 (2018)
  729--738.

\bibitem{dalca2019varreg}
A.~V. Dalca, G.~Balakrishnan, J.~Guttag, M.~Sabuncu, Unsupervised learning of
  probabilistic diffeomorphic registration for images and surfaces, Medical
  Image Analysis 57 (2019) 226--236.

\bibitem{niethammer2019metric}
M.~Niethammer, R.~Kwitt, F.-X. Vialard, Metric learning for image registration,
  in: Proceedings of the IEEE Conference on Computer Vision and Pattern
  Recognition, 2019, pp. 8463--8472.

\bibitem{wang2015predict}
Q.~Wang, M.~Kim, Y.~Shi, G.~Wu, D.~Shen, A.~D.~N. Initiative, et~al., Predict
  brain {MR} image registration via sparse learning of appearance and
  transformation, Medical Image Analysis 20~(1) (2015) 61--75.

\bibitem{jaderberg2015stn}
M.~Jaderberg, K.~Simonyan, A.~Zisserman, et~al., Spatial transformer networks,
  in: Advances in Neural Information Processing Systems, 2015, pp. 2017--2025.

\bibitem{jason2016back}
J.~Y. Jason, A.~W. Harley, K.~G. Derpanis, Back to basics: Unsupervised
  learning of optical flow via brightness constancy and motion smoothness, in:
  Proceedings of the European Conference on Computer Vision (ECCV), Springer,
  2016, pp. 3--10.

\bibitem{krebs2019learning}
J.~Krebs, H.~Delingette, B.~Mailh{\'e}, N.~Ayache, T.~Mansi, Learning a
  probabilistic model for diffeomorphic registration, IEEE Transactions on
  Medical Imaging 38~(9) (2019) 2165--2176.

\bibitem{de2019deep}
B.~D. de~Vos, F.~F. Berendsen, M.~A. Viergever, H.~Sokooti, M.~Staring,
  I.~I{\v{s}}gum, A deep learning framework for unsupervised affine and
  deformable image registration, Medical Image Analysis 52 (2019) 128--143.

\bibitem{arsigny2006log}
V.~Arsigny, O.~Commowick, X.~Pennec, N.~Ayache, A log-{E}uclidean framework for
  statistics on diffeomorphisms, in: MICCAI, Springer, 2006, pp. 924--931.

\bibitem{su2017learning}
Y.-C. Su, K.~Grauman, Learning spherical convolution for fast features from 360
  imagery, in: Advances in Neural Information Processing Systems (NIPS), 2017,
  pp. 529--539.

\bibitem{cohen2018spherical}
T.~S. Cohen, M.~Geiger, J.~K{\"o}hler, M.~Welling, Spherical {CNN}s, in:
  International Conference on Learning Representations (ICLR), 2018.

\bibitem{coors2018spherenet}
B.~Coors, A.~Paul~Condurache, A.~Geiger, Sphere{N}et: Learning spherical
  representations for detection and classification in omnidirectional images,
  in: Proceedings of the European Conference on Computer Vision (ECCV), 2018,
  pp. 518--533.

\bibitem{seong2018geometric}
S.-B. Seong, C.~Pae, H.-J. Park, Geometric convolutional neural network for
  analyzing surface-based neuroimaging data, Frontiers in Neuroinformatics 12
  (2018) 42.

\bibitem{jiang2019sphericalunet}
C.~Jiang, J.~Huang, K.~Kashinath, P.~Marcus, M.~Niessner, et~al., Spherical
  {CNN}s on unstructured grids, in: International Conference on Learning
  Representations (ICLR), 2019.

\bibitem{zhao2019sphereunet}
F.~Zhao, S.~Xia, Z.~Wu, D.~Duan, L.~Wang, W.~Lin, J.~H. Gilmore, D.~Shen,
  G.~Li, Spherical u-net on cortical surfaces: methods and applications, in:
  International Conference on Information Processing in Medical Imaging,
  Springer, 2019, pp. 855--866.

\bibitem{gopinath2019graph}
K.~Gopinath, C.~Desrosiers, H.~Lombaert, Graph convolutions on spectral
  embeddings for cortical surface parcellation, Medical Image Analysis 54
  (2019) 297--305.

\bibitem{gopinath2019adaptive}
K.~Gopinath, C.~Desrosiers, H.~Lombaert, Adaptive graph convolution pooling for
  brain surface analysis, in: International Conference on Information
  Processing in Medical Imaging, Springer, 2019, pp. 86--98.

\bibitem{henschel2020fastsurfer}
L.~Henschel, S.~Conjeti, S.~Estrada, K.~Diers, B.~Fischl, M.~Reuter,
  Fastsurfer--a fast and accurate deep learning based neuroimaging pipeline,
  NeuroImage 219.

\bibitem{polarSTN2017}
C.~Esteves, C.~Allen{-}Blanchette, X.~Zhou, K.~Daniilidis, Polar transformer
  networks, in: International Conference on Learning Representations (ICLR),
  2018.

\bibitem{tai2019equivariant}
K.~S. Tai, P.~Bailis, G.~Valiant, Equivariant transformer networks, in:
  International Conference on Machine Learning (ICML), 2019.

\bibitem{ronneberger2015u}
O.~Ronneberger, P.~Fischer, T.~Brox, U-net: Convolutional networks for
  biomedical image segmentation, in: International Conference on Medical Image
  Computing and Computer-Assisted Intervention, Springer, 2015, pp. 234--241.

\bibitem{mueller2005adni}
S.~G. Mueller, M.~W. Weiner, L.~J. Thal, R.~C. Petersen, C.~R. Jack, W.~Jagust,
  J.~Q. Trojanowski, A.~W. Toga, L.~Beckett, Ways toward an early diagnosis in
  {A}lzheimer’s disease: the {A}lzheimer’s disease neuroimaging initiative
  ({ADNI}), Alzheimer's \& Dementia 1~(1) (2005) 55--66.

\bibitem{van2001buckner}
J.~D. Van~Horn, J.~S. Grethe, P.~Kostelec, J.~B. Woodward, J.~A. Aslam, D.~Rus,
  D.~Rockmore, M.~S. Gazzaniga, The functional {M}agnetic {R}esonance {I}maging
  data center ({fMRIDC}): the challenges and rewards of large--scale databasing
  of neuroimaging studies, Philosophical Transactions of the Royal Society of
  London B: Biological Sciences 356~(1412) (2001) 1323--1339.

\bibitem{desikan2006automated}
R.~S. Desikan, F.~S{\'e}gonne, B.~Fischl, B.~T. Quinn, B.~C. Dickerson,
  D.~Blacker, R.~L. Buckner, A.~M. Dale, R.~P. Maguire, B.~T. Hyman, et~al., An
  automated labeling system for subdividing the human cerebral cortex on {MRI}
  scans into gyral based regions of interest, NeuroImage 31~(3) (2006)
  968--980.

\bibitem{van2013hcp}
D.~C. Van~Essen, S.~M. Smith, D.~M. Barch, T.~E. Behrens, E.~Yacoub,
  K.~Ugurbil, W.-M.~H. Consortium, et~al., The {WU-Minn} human connectome
  project: an overview, NeuroImage 80 (2013) 62--79.

\bibitem{glasser2013minimal}
M.~F. Glasser, S.~N. Sotiropoulos, J.~A. Wilson, T.~S. Coalson, B.~Fischl,
  J.~L. Andersson, J.~Xu, S.~Jbabdi, M.~Webster, J.~R. Polimeni, et~al., The
  minimal preprocessing pipelines for the {Human Connectome Project},
  NeuroImage 80 (2013) 105--124.

\bibitem{barch2013function}
D.~M. Barch, G.~C. Burgess, M.~P. Harms, S.~E. Petersen, B.~L. Schlaggar,
  M.~Corbetta, M.~F. Glasser, S.~Curtiss, S.~Dixit, C.~Feldt, et~al., Function
  in the human connectome: task-{fMRI} and individual differences in behavior,
  NeuroImage 80 (2013) 169--189.

\bibitem{dice1945measures}
L.~R. Dice, Measures of the amount of ecologic association between species,
  Ecology 26~(3) (1945) 297--302.

\bibitem{khasanova2017weight}
R.~Khasanova, P.~Frossard, Graph-based classification of omnidirectional
  images, in: Proceedings of the IEEE International Conference on Computer
  Vision (ICCV), 2017, pp. 869--878.

\bibitem{hinds2008accurate}
O.~P. Hinds, N.~Rajendran, J.~R. Polimeni, J.~C. Augustinack, G.~Wiggins, L.~L.
  Wald, H.~D. Rosas, A.~Potthast, E.~L. Schwartz, B.~Fischl, Accurate
  prediction of v1 location from cortical folds in a surface coordinate system,
  NeuroImage 39~(4) (2008) 1585--1599.

\bibitem{glasser2011mapping}
M.~F. Glasser, D.~C. Van~Essen, Mapping human cortical areas in vivo based on
  myelin content as revealed by {T1-} and {T2-weighted MRI}, Journal of
  Neuroscience 31~(32) (2011) 11597--11616.

\bibitem{tong2017functional}
T.~Tong, I.~Aganj, T.~Ge, J.~R. Polimeni, B.~Fischl, Functional density and
  edge maps: Characterizing functional architecture in individuals and
  improving cross-subject registration, NeuroImage 158 (2017) 346--355.

\end{thebibliography}

\end{document}